\documentclass[onecolumn]{aa}
\usepackage[varg]{txfonts}
\usepackage{amsmath,amssymb}
\usepackage[colorlinks=true,linkcolor=blue,citecolor=blue,filecolor=blue,urlcolor=blue]{hyperref}
\usepackage{siunitx}

\usepackage{tikz}
\usepackage{tikz-3dplot}

\begin{document}
\title{Nutation-orbit resonances: The origin of the chaotic rotation of Hyperion and the barrel instability}

\author{Max Goldberg\inst{\ref{inst1},\ref{inst2}} \and Konstantin Batygin\inst{\ref{inst3}}}

\institute{
Laboratoire Lagrange, Université Côte d'Azur, CNRS, Observatoire de la Côte d'Azur, Boulevard de l'Observatoire, 06300 Nice Cedex 4, France\label{inst1}
\and 
Department of Astronomy, California Institute of Technology, Pasadena, CA 91125, USA\label{inst2}
\and 
Division of Geological and Planetary Science, California Institute of Technology, Pasadena, CA 91125, USA\label{inst3}}
\date{}

\abstract{While numerous planetary and asteroid satellites show evidence for non-trivial rotation states, none are as emblematic as Hyperion, which has long been held as the most striking example of chaotic spin-orbit evolution in the Solar System. Nevertheless, an analytically tractable theory of the full 3D spin-orbit dynamics of Hyperion has not been developed.
We derive the Hamiltonian for a spinning axisymmetric satellite in the gravitational potential of a planet without assuming planar or principal axis rotation and without averaging over the spin period. Using this model, we demonstrate the emergence of resonances between the nutation and orbital frequencies that act as the primary drivers of the spin dynamics. This analysis reveals that, contrary to long-held belief, Hyperion is not tumbling chaotically. Instead, it lies near or in a nutation-orbit resonance that is first-order in eccentricity, allowing it to rotate quasi-regularly. The most reliable observations are consistent with either nonchaotic motion or chaos that is orders of magnitude smaller than originally claimed. A separate phenomenon, the so-called barrel instability, is shown to be related to a different set of nutation-orbit resonances that generalize the planar spin-orbit resonances. Finally, we show that changes in spin states over long timescales are best understood by considering chaotic diffusion of quasi-conserved quantities.}

\titlerunning{Nutation-orbit resonances}
\authorrunning{Goldberg \& Batygin}
\maketitle

\section{Introduction}
Nearly all regular satellites in the Solar System rotate synchronously with their mean orbital motion as a consequence of tidal evolution. 
In a landmark paper, \cite{Wisdom1984} demonstrated that Hyperion is an exception to this rule: due to its unusually large eccentricity and asphericity, a large chaotic sea surrounds the 1:2, 1:1, 3:2, and 2:1 planar spin-orbit resonances, facilitating chaotic spin evolution.
Furthermore, the chaotic sea and most of the resonances are attitude unstable, so that the obliquity of Hyperion would immediately depart from zero if it began there.
Their work helped set off a flurry of research that elevated chaos from mathematical curiosity to concrete reality in the Solar System \citep{Wisdom1987,Laskar1989}.

A concrete prediction of \cite{Wisdom1984} was that Hyperion would be found in a chaotically tumbling state with a spin rate of approximately 0.5 to 2 times its orbital frequency.
However, measurements of Hyperion's spin state show a rather different configuration. 
Hyperion rotates rapidly, more than four times in a single orbit, and its rotation axis is closely aligned to its longest axis \citep{Thomas1995}. 
Observations made during spacecraft flybys separated by more than twenty years show a strikingly consistent spin state, suggesting little or no evolution \citep{Harbison2011}.
Even so, numerical analyses repeatedly find that rotation of Hyperion is in fact chaotic with a Lyapunov timescale of a few orbits \citep{Black1995,Harbison2011}. 
These seemingly contradictory results have not been satisfactorily explained in the literature.

Beyond the specific case of Hyperion, recent investigations of the spin evolution of planetary satellites and asteroid binaries over a large parameter space have uncovered a rich assortment of theoretical spin configurations that appear to be long-lived in addition to the expected synchronous, chaotic tumbling, and rapid rotation states. 
These include apparent irregular rotation that nevertheless shows a preference for particular orientations \citep{MelNikov2000,Cuk2021}, resonances between precession and orbital frequencies \citep{Benettin2008,Agrusa2021}, and complex alternation between high obliquity states and tumbling \citep{Quillen2017}.
While dynamically interesting in their own right, these exotic spin states also have major implications for the efficiency of tidal spin down and the BYORP effect \citep{Cuk2021,Quillen2020,Quillen2022}.

Our objective is to develop a theory of rotation that can describe and classify these spin configurations.
We develop an analytical model of the rotation of an axisymmetric ellipsoid in a Newtonian gravitational potential that does not assume a particular orientation, obliquity, or circular orbit. 
This model exhibits a set of resonances between the nutation frequency, or ``wobble,'' and the orbital motion, which have not been considered previously in the literature.
Observations of the rotation of Hyperion suggest that it is in, or near, one of these nutation-orbit resonances.
We also demonstrate that chaos on short timescales can arise from the overlap of these resonances; a straightforward time-dependent one degree-of-freedom model recovers the chaotic behavior of Hyperion while revealing two integrals of motion.
Additionally, a separate set of nutation-orbit resonances generalizes the well-known spin-orbit resonances but allows for non-principal axis (NPA) rotation and obliquity.
We show that the barrel instability is a consequence of capture into one of these resonances.
Finally, we argue that long term variations in the spin state, observed in numerical simulations, are due to the slow diffusion of quasi-integrals of motion.

Previous attempts in the literature to analytically study NPA rotation have generally followed one of two approaches.
Many authors have demonstrated that for certain parameters, synchronous or near-synchronous spin states are attitude unstable
\citep{MelNikov1998,Melnikov2008,Gaitanas2024,Tan2023}. This approach dates to \cite{Wisdom1984}, who showed that, even though a synchronous spin-orbit resonance exists for Hyperion in the planar problem, any small obliquity would quickly grow and Hyperion would immediately leave the synchronous state. 
While this technique can properly answer the question about whether a truly synchronous state is possible, it can only reveal local behavior around the planar fixed point and misses the existence of other fixed points and quasi-periodic orbits in other regions of parameter space.

Alternatively, some authors have developed general models of NPA rotation and spin-orbit coupling \citep{Kinoshita1972,Ferrandiz1989,Boue2009,Crespo2018}. 
These studies have revealed the existence of additional equilibria and conserved quantities.
However, each of these works reduces the complexity of the problem by averaging over a fast angle, typically the nutation angle. 
As such, this procedure has the side effect of removing resonances and chaotic behavior associated with the angle averaged over. 
Therefore, to study these dynamics in their full detail while remaining analytically tractable, we refrain from averaging over rotation angles at the cost of additional complexity.

\section{Model of rotational dynamics}
\label{sec:model}
We begin by deriving the Hamiltonian for a spinning rigid axisymmetric ellipsoid in an arbitrary Keplerian orbit. The inertial laboratory frame, which we will relate to the orbit later, is defined by the orthonormal basis vectors $\mathbf{\hat{x}}$, $\mathbf{\hat{y}}$, and $\mathbf{\hat{z}}$.
\subsection{Free rotation}
\begin{figure}
    \centering
    \includegraphics[width=0.5\linewidth]{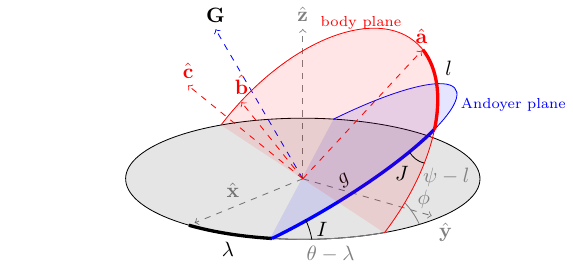}
    \caption{Definitions of the Andoyer coordinates in terms of the laboratory $xy$-plane, Andoyer plane, and body equatorial $ab$-plane. Also shown are the standard Euler angles $(\theta,\phi,\psi)$ in the ZXZ convention. The Andoyer actions are related by $\cos{I} = \Lambda/G$ and $\cos{J}= L/G$.}
    \label{fig:andoyer}
\end{figure}

The ellipsoid is taken to have fixed principal moments of inertia $A$, $B$, and $C$ along its $\mathbf{\hat{a}}$, $\mathbf{\hat{b}}$, and $\mathbf{\hat{c}}$ principal axes, which define the body frame. The components of the angular velocity vector $\pmb{\omega}$ along these axes are $(\omega_a, \omega_b, \omega_c)$. The rotational kinetic energy of the spinning ellipsoid, $((A\omega_a)^2 + (B\omega_b^2) + (C \omega_c)^2)/2$, is clearly conserved, but these momenta are not conjugate to convenient angles. 
To write the Hamiltonian, we instead use the Andoyer canonical coordinates ($G$, $g$, $\Lambda$, $\lambda$, $L$, $l$) of \cite{Deprit1967}. 
These coordinates make use of an intermediate plane normal to the spin angular momentum vector $\mathbf{G}$ whose $x$-axis is the intersection of the intermediate plane with the inertial $xy$-plane.
Then, $\Lambda = \mathbf{G}\cdot \mathbf{\hat{z}}$ is the projection of $\mathbf{G}$ on the inertial $z$-axis, and $\lambda$ is the longitude of the intersection of the intermediate plane with the inertial $xy$-plane. 
Furthermore, $G=|\mathbf{G}|$ is the norm of the spin angular momentum, and $g$ is the argument of the intersection of the equatorial, or $ab$, plane of the body with the intermediate plane. 
Finally, $L=\mathbf{G}\cdot \mathbf{\hat{c}}=C\omega_c$ is the projection of $\mathbf{G}$ on the $c$-axis of the body, and $l$ is the argument of the $a$-axis of the body in the equatorial plane.
A diagram of these frames and coordinates is shown in Fig.~\ref{fig:andoyer}, and details of how to transform between these coordinates and other typical ways of describing rotation are provided in Appendix \ref{sec:A1}.

The Hamiltonian in Andoyer canonical coordinates of a freely rotating ellipsoid is \citep{Deprit1967}
\begin{equation}
    \mathcal{H}_\textrm{free} = \frac{1}{2}\left(\frac{\cos^2{l}}{B} + \frac{\sin^2{l}}{A}\right)\left(G^2-L^2\right) + \frac{1}{2C}L^2.
\end{equation} 
We assume the ellipsoid is axisymmetric, meaning that two of its moments of inertia are equal, by setting $B=A$. Then, the Hamiltonian is simply
\begin{equation}
    \mathcal{H}_\textrm{free,axi} = \frac{G^2}{2A} + L^2 \left(\frac{1}{2C} - \frac{1}{2A}\right).
\end{equation} 
This Hamiltonian is independent of $\lambda$, $g$, $l$, and $\Lambda$, and thus $\Lambda$, $G$, $L$, and $\lambda$ are conserved. The remaining coordinates can be trivially integrated as the remaining derivatives $\dot{g} = \partial{\mathcal{H}_\textrm{free,axi}}/\partial{G} = G/A$ and $\dot{l} = \partial{\mathcal{H}_\textrm{free,axi}}/\partial{L} = L(1/C-1/A)$ are constant in time.

It is instructive to pause momentarily for interpretation of these expressions. In the principal axis (body) frame, the ellipsoid $c$ axis moves around the angular momentum vector at the angular frequency $\dot{l} = L(1/C-1/A)$. 
For Hyperion, this corresponds to a period of $\approx \qty{15}{d}$, in close agreement with the period of the ``body frame wobble'' found by \cite{Black1995}.
Furthermore, in the frame of the intermediate plane, the $c$ axis evolves around the normal to the intermediate plane at the rate $\dot{g} = G/A$. 
For typical, nearly spherical bodies ($A\approx C$), this frequency is comparable to the total angular velocity.
In the case of Hyperion, we find the period of $g$ to be $\approx \qty{7}{d}$, again in close agreement with the inertial wobble frequency found by \cite{Black1995}. 
In the near-principal axis rotation typical of planets, $L \approx G$, and both $g$ and $l$ become ill-defined. In that case, the total spin rate around the $c$-axis is $\dot{g} + \dot{l} \approx G/C \approx \omega_c$. 

For clarity, we will use the term ``nutation'' to refer to evolution of $g$ and $l$, which characterize the orientation of the body relative to the angular momentum.
We will reserve the term ``precession'' solely to describe the change of the angular momentum vector in inertial space, $\dot{\lambda}$, which is absent in the torque-free case.
We also emphasize that we do not assume that $A\leq B\leq C$. Instead, the choice of assigning the $a$, $b$, and $c$ axes is determined by the axisymmetry assumption that $A=B$. Therefore, in this context, prolate bodies correspond to $A>C$, while oblate bodies have $A<C$. 

\subsection{Tidal potential}
We now calculate the gravitational potential $\Phi_\mathcal{G}$ on the satellite due to the tidal torque as it orbits the primary in a fixed Keplerian orbit.
We can assume without loss of generality that the orbital plane lies on the inertial $xy$ plane, and the pericenter of the orbit lies on the $x$-axis.
According to MacCullagh's formula, the potential on the satellite exerted by the primary, to second order in the ratio of satellite radius to the orbital radius, is
\begin{equation}
    \Phi_\mathcal{G} = -\frac{1}{2r^3}\left((B+C-2A)\alpha^2 + (C+A-2B)\beta^2 + (A+B-2C)\gamma^2\right)
\end{equation}
where $\alpha$, $\beta$, and $\gamma$ are the direction cosines between the principal axes and the direction of the primary, and $r$ is the primary-satellite distance \citep{Murray1999}. Then, the Hamiltonian describing the spin evolution of a triaxial ellipsoid in a fixed Keplerian orbit is
\begin{equation}
    \mathcal{H} = \mathcal{H}_\textrm{free} + \Phi_\mathcal{G}
\end{equation}
which we will not write explicitly in terms of the Andoyer actions and angles for brevity.

Using the axisymmetry assumption and the fact that $\alpha^2+\beta^2+\gamma^2=1$, the potential simplifies to
\begin{equation}
    \Phi_\mathrm{axi}=-\frac{1}{2r^3}(A-C)(1-3\gamma^2).
\end{equation}
Furthermore, in Andoyer coordinates, 
\begin{equation}
\gamma = \sin J \cos(f-\lambda)\sin g - (\cos J \sin I + \sin J \cos I \cos g)\sin(f-\lambda)
\label{eq:gamma}
\end{equation}
where $\cos I = \Lambda/G$ and $\cos J = L/G$, and $f$ is the true anomaly \citep{Lara2010}. After substitution and dropping of constant terms, the Hamiltonian of the axisymmetric spinning ellipsoid in a fixed Keplerian orbit is
\begin{align}
    \hat{\mathcal{H}} = \mathcal{H}_\textrm{free,axi} + \Phi_\mathrm{axi} =  \frac{G^2}{2A} + L^2\left(\frac{1}{2C}
    - \frac{1}{2A}\right) + \frac{3}{16r^3}(A-C)\left( - 2c^2_J - 2c^2_I + 6c^2_J c^2_I \right. \nonumber \\
    - 2 s_{2J} s_{2I} \cos(g) + 2s^2_J s^2_I\cos(2g) - 2\left(1 - 3c^2_J - c^2_I + 3c^2_J c^2_I\right) \cos(2\lambda - 2f) \nonumber \\
    + s^2_J\left(1 - c_I\right)^2\cos(2g-2\lambda + 2f) + s^2_J\left(1 + c_I\right)^2\cos(2g+2\lambda-2f) \nonumber \\
    + 2 s_{2J} s_I\left(-1 + c_I\right)\cos(g-2\lambda + 2f) + \left. 2s_{2J} s_I\left(1 + c_I\right)\cos(g + 2\lambda - 2f) \right)
\end{align}
where we have written $c_x$ and $s_x$ for $\cos(x)$ and $\sin(x)$, respectively.

One additional simplification is possible. We scale each of the actions by $A$, making the substitutions $\tilde{G}=G/A$, $\tilde{L}=L/A$, and $\tilde{\Lambda}=\Lambda/A$. To maintain symplecticity (i.e., to avoid modifying the time unit), we also scale the Hamiltonian by the same factor, $\tilde{\mathcal{H}}=\hat{\mathcal{H}}/A$. This has the effect of reducing the dependence on the moments of inertia to a single quantity, which we define to be $\rho = 3(A-C)/A$. Henceforth we will drop the tildes on the actions for clarity, but we refer to the scaled actions unless explicitly indicated otherwise. Then, the Hamiltonian, written explicitly in terms of the actions, is
\newcommand{\LoverG}{\frac{L}{G}}
\newcommand{\LamoverG}{\frac{\Lambda}{G}}
\newcommand{\LoverGsq}{\frac{L^2}{G^2}}
\newcommand{\LamoverGsq}{\frac{\Lambda^2}{G^2}}
\begin{align}
    \tilde{\mathcal{H}} = \frac{G^2}{2} + \frac{L^2}{2}\frac{\rho}{3-\rho} + \frac{1}{16r^3}\rho\left( - 2\LoverGsq - 2\LamoverGsq + 6\LoverGsq\LamoverGsq \right. \nonumber \\
    - 8 \LoverG\sqrt{1 - \LoverGsq} \LamoverG\sqrt{1 - \LamoverGsq} \cos(g) + 2\left(1 - \LoverGsq\right)\left(1 - \frac{\Lambda^2}{G^2}\right)\cos(2g) \nonumber \\
    - 2\left(1 - 3\LoverGsq - \LamoverGsq + 3\LoverGsq\LamoverGsq\right) \cos(2\lambda - 2f) \nonumber \\
    + \left(1- \LoverGsq\right)\left(1 - \LamoverG\right)^2\cos(2g-2\lambda+2f) \nonumber \\
    + \left(1- \LoverGsq\right)\left(1 + \LamoverG\right)^2\cos(2g+2\lambda-2f) \nonumber \\
    + 4\left(\LoverG\sqrt{1-\LoverGsq}\sqrt{1-\LamoverGsq}\left(-1 + \LamoverG\right)\right)\cos(g - 2\lambda + 2f) \nonumber \\
    + \left. 4\left(\LoverG\sqrt{1 - \LoverGsq}\sqrt{1 - \LamoverGsq}\left(1 + \LamoverG\right)\right)\cos(g + 2\lambda - 2f) \right).
\end{align}
Hamiltonian $\tilde{\mathcal{H}}$ is no longer independent of $g$, $\lambda$, or $\Lambda$, and thus their conjugate pairs, which were conserved in free rotation, can vary. However, it is cyclic in $l$ and thus $L$ remains conserved. It is also explicitly time-dependent through $r$ and $f$, both of which vary nonlinearly in time if the orbit is eccentric. 

This model is indeed a generalization of the standard spin-orbit prescription of \cite{Goldreich1966}, who assume that the spin vector is aligned to both the orbit normal and to a principal axis. That is, the ellipsoid rotates solely around the body $a$-axis, which is aligned to the inertial $z$-axis, so that $L=0$ and $\Lambda=G$. In such a state, $\lambda$ is ill-defined and can be set arbitrarily to 0. Then, the Hamiltonian becomes
\begin{equation}
    \mathcal{H}_\textrm{planar} = \frac{G^2}{2} + \frac{\rho}{4r^3}\cos(2g-2f),
\end{equation}
which has the pendulum-like equation of motion
\begin{equation}
    \ddot{g} = \frac{d}{dt} \left(\frac{\partial \mathcal{H}_\textrm{planar}}{\partial G}\right) = \dot{G} = -\frac{\partial\mathcal{H}_\textrm{planar}}{\partial g} = \frac{\rho}{2r^3}\sin(2g-2f),
\end{equation}
equivalent to that derived by \cite{Goldreich1966} under the axisymmetry assumption. Our generalized model therefore includes these previously studied dynamics but also allows for a much broader set of behaviors associated with a spin axis misaligned to both the orbital plane and the body principal axes.

\section{Rotation of Hyperion}
The shape of Hyperion resembles a nearly axisymmetric ellipsoid. 
\cite{Thomas1995} estimated the principal moments of inertia from the Voyager 2 shape model assuming a homogeneous interior. Converting their results to our convention of $A\approx B$ via a permutation of the principal axes (see Appendix \ref{sec:A1}), they derived $A_H=0.459$, $B_H=0.519$, and $C_H=0.323$, normalized by $M_\textrm{H}\langle R_\textrm{H}\rangle^2$ where $M_\textrm{H}$ is the mass of Hyperion and $\langle R_\textrm{H} \rangle$ is its mean radius. \cite{Harbison2011} performed a similar analysis using the Cassini shape model of \cite{Thomas2007} and found $A_H=0.474$, $B_H=0.542$, $C_H=0.314$. 
To conform to the axisymmetry assumption, we take $A=(A_H+B_H)/2$ and $C=C_H$, thus obtaining $\rho=1.02$ and $\rho=1.15$ for the \cite{Thomas1995} and \cite{Harbison2011} measurements, respectively.

Hyperion's orbit around Saturn has a period of \qty{21.2}{d}, corresponding to a semi-major axis of about 25 Saturn radii.
Its eccentricity varies from approximately 0.08 to 0.12 with a period of 20 years because it is locked in a 4:3 mean-motion resonance with Titan \citep{Duriez1997}. 
This eccentricity, uniquely large among the regular satellites of the Solar System \citep{Peale1999}, is believed to have originated in the mean-motion resonant interaction with an outwardly migrating Titan \citep{Colombo1974,Cuk2013,Goldberg2024}.
\begin{table}[]
    \caption{Observed orbit and spin states of Hyperion during the single Voyager 2 and the three Cassini flybys.}
    \centering
    \begin{tabular}{c|c|c|c|c|c|c|c|c|c}
         Date & $e$ & $M$ & $\theta$ & $\phi$ & $\psi$ & $\omega_A/|\omega|$ & $\omega_B/|\omega|$ & $\omega_C/|\omega|$ & $|\omega|$ \\   
         \hline
         \multicolumn{10}{c}{Uncorrected for body axis-principal axis misalignment}\\
         \hline
         1981-08-23 & 0.124 & 0.349 & 5.676 & 0.965 & 3.448 & 0.211 & -0.019 & 0.977 & 4.225 \\
         2005-06-10 & 0.115 & 5.149 & 1.713 & 1.148 & 1.508 & 0.067 & 0.451 & 0.890 & 4.433 \\
         2005-08-16 & 0.115 & 5.969 & 0.559 & 0.661 & 0.587 & 0.162 & 0.389 & 0.907 & 4.255 \\
         2005-09-25 & 0.113 & 5.288 & 5.734 & 0.134 & 0.705 & 0.133 & 0.411 & 0.902 & 4.255 \\
         \hline
         \multicolumn{10}{c}{Corrected with estimate of body axis-principal axis misalignment} \\
         \hline
         2005-06-10 & 0.115 & 5.149 & 2.055 & 0.963 & 2.210 & -0.010 & -0.182 & 0.983 & 4.433 \\
         2005-08-16 & 0.115 & 5.969 & 1.025 & 0.819 & 1.112 & -0.166 & 0.009 & 0.986 & 4.255 \\
         2005-09-25 & 0.113 & 5.288 & 0.527 & 0.393 & 0.525 & -0.095 & 0.157 & 0.983 & 4.255
    \end{tabular}
    \tablefoot{Data reported by \cite{Black1995} and \cite{Harbison2011}. The Euler angles $\theta$, $\phi$, and $\psi$ use the ZXZ convention.}
    \label{tab:cassini}
\end{table}
\begin{table}[]
    \caption{Andoyer coordinates of the observed rotation states from Table \ref{tab:cassini}, including the rescaled actions.}
    \centering
    \begin{tabular}{c|c|c|c|c|c|c|c|c|c}
         Date & $G$ & $\Lambda$ & $L$ & $\tilde{G}$ & $\tilde{\Lambda}$ & $\tilde{L}$ & $g$ & $\lambda$ & $l$ \\
         \hline
         \multicolumn{10}{c}{Uncorrected for body axis-principal axis misalignment}\\
         \hline
         1981-08-23 & 1.397 & 0.597 & 1.335 & 2.855 & 1.219 & 2.728 & 2.221 & 5.688 & 1.673 \\
         2005-06-10 & 1.652 & 0.776 & 1.239 & 3.252 & 1.528 & 2.439 & 1.671 & 0.827 & 0.129 \\
         2005-08-16 & 1.543 & 0.596 & 1.212 & 3.037 & 1.174 & 2.385 & 2.589 & 1.997 & 0.349 \\
         2005-09-25 & 1.557 & 1.068 & 1.205 & 3.064 & 2.102 & 2.372 & 3.191 & 2.673 & 0.276 \\
         \hline
         \multicolumn{10}{c}{Corrected with estimate of body axis-principal axis misalignment} \\
         \hline
         2005-06-10 & 1.437 & 0.599 & 1.369 & 2.828 & 1.178 & 2.694 & 4.032 & 1.224 & 3.192 \\
         2005-08-16 & 1.360 & 0.683 & 1.318 & 2.676 & 1.344 & 2.594 & 3.543 & 2.162 & 4.777 \\
         2005-09-25 & 1.376 & 1.056 & 1.314 & 2.708 & 2.080 & 2.586 & 3.118 & 2.598 & 5.796
    \end{tabular}
    \label{tab:andoyer}
\end{table}

The rotation states observed during the Voyager 2 flyby in 1981 \citep{Thomas1995,Black1995} and Cassini flybys in 2005 \citep{Thomas2007,Harbison2011} are summarized in Table \ref{tab:cassini} and the values in Andoyer coordinates are given in Table \ref{tab:andoyer}. In all cases, rotation predominantly occurs on the long axis. 
The components of rotation on the short body axes vary, as does the projection of the spin vector in inertial space. 
The total spin frequency is remarkably constant at approximately 4.2 times the mean motion, suggesting the existence of a symmetry that is not immediately evident in $\tilde{\mathcal{H}}$.

The full, three degree-of-freedom dynamics of even the axisymmetric Hamiltonian $\tilde{\mathcal{H}}$ is too complex to analyze in general. Therefore, we must make simplifying assumptions relevant for the problem at hand. 
A conventional approach in celestial mechanics is to average over circulating harmonics, so the first point to consider is which of the seven cosine harmonics must be retained. Harmonics whose argument varies rapidly can be averaged over without significantly affecting the overall dynamics. 
The rate of change of each harmonic's argument can be estimated by assuming that the evolution of the coordinates is dominated by the kinetic term in the Hamiltonian. For the parameters of Hyperion, and considering only the kinetic term, we have that $\dot{g} \approx \tilde{G} \approx 3$, $|\dot{\lambda}| \ll 1$, and $\dot{f}\approx 1$. This reasoning suggests that the final harmonic, of argument $\phi \equiv g + 2\lambda - 2f$, which will have $\dot{\phi} \approx 1$, is likely to be most dynamically important in the context of Hyperion's current rotation state.

\subsection{Single harmonic}
\label{sec:sing}
Therefore, as a simple first approximation we retain only the final harmonic of $\tilde{\mathcal{H}}$, which has the argument $g+2\lambda - 2f$. This reduced Hamiltonian $\hat{\mathcal{H}}^1$ depends on the coordinates only through the single linear combination $g+2\lambda$. To eliminate a degree of freedom, we define the canonical transformation $g'=g+2\lambda$, $\Lambda'=\Lambda - 2G$, and $G'=G$ and $\lambda'=\lambda$ are left unchanged, so we drop their primes. We then obtain
\begin{align}
    \hat{\mathcal{H}}^1 = \frac{G^2}{2} + \frac{\rho}{8r^3}\left( - \frac{L^2}{G^2} - \frac{(\Lambda' + 2G)^2}{G^2} + 3\frac{L^2}{G^2} \frac{(\Lambda' + 2G)^2}{G^2}\right. \nonumber\\
    + \left. 2\frac{L}{G} \sqrt{1 - \frac{L^2}{G^2}} \sqrt{1 - \frac{(\Lambda' + 2G)^2}{G^2}} \left(\frac{\Lambda'}{G} - 1\right)\cos(g' - 2f) \right).
    \label{eq:Honeharm}
\end{align}
Hamiltonian $\hat{\mathcal{H}}^1$ is now cyclic in $\lambda$ and therefore $\Lambda'=\Lambda - 2G$ is a conserved quantity.

Although superficially quite different because of the complex dependence on the actions, $\hat{\mathcal{H}}^1$ is very similar to the standard planar spin-orbit model of \cite{Goldreich1966}. Following their approach, we expand $r$ and $f$ in powers of eccentricity to obtain the analogue of spin-orbit resonances \citep{Goldreich1966,Wisdom1984}. Retaining terms up to order $e^2$ and averaging over cosines that depend only on $M$, we obtain
\begin{align}
    \hat{\mathcal{H}}^1 \approx \frac{G^2}{2} + \frac{\rho}{8}\left(\left(1 + \frac{3e^2}{2}\right) \left( - \frac{L^2}{G^2} - \frac{(\Lambda' + 2G)^2}{G^2} + 3\frac{L^2}{G^2} \frac{(\Lambda' + 2G)^2}{G^2}\right)\right. \nonumber \\
    + 2\left(\frac{L}{G} \sqrt{1 - \frac{L^2}{G^2}} \sqrt{1 - \frac{(\Lambda' + 2G)^2}{G^2}} \left(\frac{\Lambda'}{G} - 1\right)\right) \left(\left(1 - \frac{5e^2}{2}\right) \cos(g' - 2M) \right. \nonumber \\
    + \frac{7e}{2} \cos(g' - 3M) \nonumber \\ 
    - \frac{e}{2} \cos(g' - M) \nonumber \\ 
    + \left. \frac{17e^2}{2} \cos(g' - 4M) \right)\Big) + O(e^3)
\end{align}
where $M=t + M_0$ is the mean anomaly. 
In general, resonances exist when $\dot{g}' = \dot{g} + 2\dot{\lambda} \simeq p$, where $p$ is a nonzero integer. We term these resonances $p$:1 ``nutation-orbit resonances.'' 
The leading term of the coefficient multiplying each resonance scales as $e^{|p-2|}$.
Intuitively, nutation-orbit resonances represent spin states in which the nutation of the spinning body around its angular momentum vector occurs at multiple of the orbital frequency, so that the body returns to its original orientation after $1/p$ orbits.
The presence of the $2\lambda$ term is required by the rotation symmetry of the problem, equivalent to the d'Alembert rules in the expansion of the disturbing function in celestial mechanics.\footnote{Note that in our coordinate system, the pericenter direction of Hyperion is set to the positive x-axis, so that $\varpi=0$. This assumption can be relaxed by replacing $f$ with $f+\varpi$ throughout.}

A relatively accurate model of Hyperion's dynamics can be obtained by keeping only the 3:1 resonant harmonic and neglecting (i.e., averaging over) the other cosine terms. To remove the explicit time dependence, we extend the phase space by adding to the Hamiltonian a momentum $T$ conjugate to $M$, and then perform the canonical substitution $\phi\equiv g' - 3M$ conjugate to $G$, and $\Theta\equiv 3G+T$ conjugate to $M$. The transformed Hamiltonian,
\begin{align}
    \mathcal{H}_3^1 = \frac{G^2}{2} - 3G + \frac{\rho}{8}\left(\left(1 +  \frac{3e^2}{2}\right) \left( - \frac{L^2}{G^2} - \left(2 + \frac{\Lambda'}{G}\right)^2 + 3\frac{L^2}{G^2} \left(2 + \frac{\Lambda'}{G}\right)^2\right) \right. \nonumber \\ 
    +\left. 7e \frac{L}{G} \sqrt{1 - \frac{L^2}{G^2}} \sqrt{1 - \left(2 + \frac{\Lambda'}{G}\right)^2} \left(\frac{\Lambda'}{G} - 1\right) \cos\phi\right) + \Theta,
\end{align}
being cyclic in $M$, has only one degree of freedom and is therefore integrable and nonchaotic. 
Nevertheless, $\mathcal{H}_3^1$ is sufficient to capture much of the current spin dynamics of Hyperion. 

\begin{figure}
    \centering
    \includegraphics[width=0.98\linewidth]{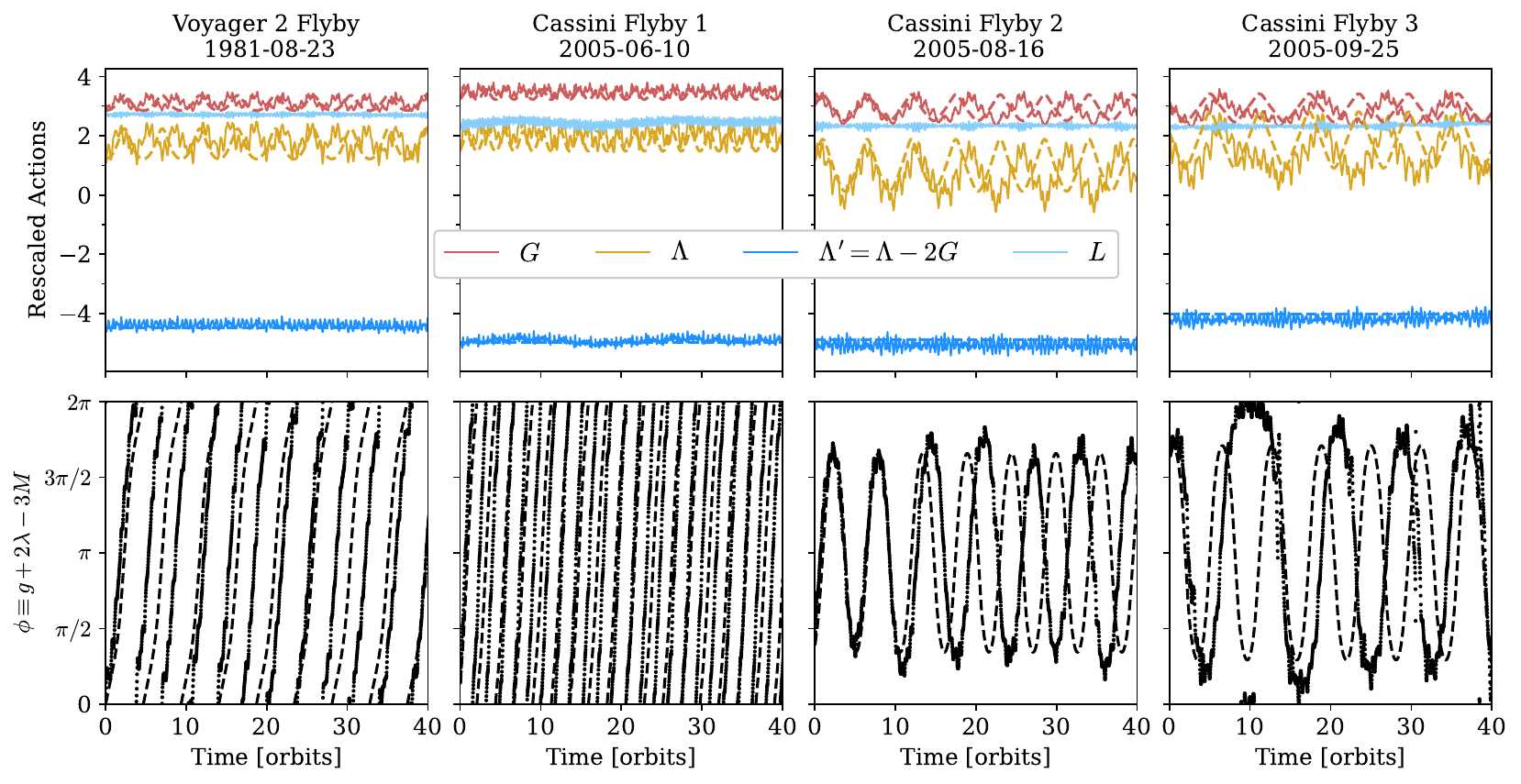}
    \caption{Time evolution of the rotation state of Hyperion in scaled Andoyer actions (top row) and the 3:1 nutation-orbit resonant angle (bottom row) following four spacecraft flybys, each of which occur at $t=0$. Solid lines show the result of the full rotation model $\mathcal{H}$, which includes nonaxisymmetry of Hyperion. The solution of the integrable one degree-of-freedom (i.e., non-chaotic) Hamiltonian $\mathcal{H}^1_3$ is shown in dashed lines, integrated from the same initial conditions. The actions $L$ and $\Lambda'$ are conserved in $\mathcal{H}^1_3$.}
    \label{fig:actions}
\end{figure}

Figure \ref{fig:actions} shows the evolution of the Andoyer actions and the resonant angle $\phi$ starting from the states observed during the Voyager 2 and Cassini flybys (Tables \ref{tab:cassini} \& \ref{tab:andoyer}). 
We numerically integrated the full Euler rigid body equations in quaternion coordinates using the same method as \cite{Goldberg2024}, incorporating the three unique moments of inertia estimated by \cite{Black1995} (for the Voyager 2 flyby) and \cite{Harbison2011} for the Cassini observations.
We also numerically integrated the equations of motion of $\mathcal{H}_3^1$, shown as dashed lines. 
We find that $\phi$ is circulating in the state observed during the Voyager 2 encounter and the first Cassini flyby (left two column), while it is librating with large amplitude in the other two Cassini 
flybys (center and right column). 
The integrable model correctly captures both libration and circulation of $\phi$, depending on initial conditions.
Furthermore, $L$ and $\Lambda'$, which are conserved in $\mathcal{H}_3^1$, vary only slightly in $\mathcal{H}$.

Several features of the complete dynamics are not present in the integrable model. 
For one, $\mathcal{H}_3^1$ overestimates the libration frequency for the second and third Cassini flyby. These solutions lie very close to the resonant separatrix, where the libration frequency depends strongly on the exact location of the separatrix relative to the state of the system. The true separatrix ``breathes'' due to perturbations from terms that we have neglected, so the solution is not periodic, and indeed can alternate between libration and circulation over timescales of tens to hundreds of orbits.
Furthermore, there are high-frequency, low amplitude oscillations in the actions. 
The variations in $L$ are associated with the $\sim 10\%$ difference between the $A$ and $B$ moments of inertia, which changes the angular velocity on the long axis during one rotation cycle \citep{Deprit1967}.
Changes in $\Lambda'$ are primarily due the presence of fast terms in the remaining six harmonics, which could be accounted for in a near-identity canonical transformation in a more elaborate calculation.

It is important to note that the evolution following the three Cassini flybys shown in Fig.~\ref{fig:actions} overlap. Each flyby is separated by just $\sim 2$ orbits and therefore the states reported by \cite{Harbison2011} are not consistent with each other. This cannot be due entirely to chaotic evolution: $\Lambda'$ is much more different between Flyby 2 and Flyby 3 than it can evolve over even hundreds of orbits. This inconsistency was noted by \cite{Harbison2011}, who suggested that the low density of Hyperion implies it must have significant voidspace. If the voids in Hyperion are not homogeneous, the moments of inertia estimated from the shape model will be inaccurate. More importantly, the true principal axes will not be parallel to the ellipsoid axes, and the reported values of $\omega_a$, $\omega_b$, $\omega_c$ will be inaccurate because they represent the projection of $\pmb{\omega}$ (which is well-measured in inertial space) onto the wrong axes.

\begin{figure}
    \centering
    \includegraphics[width=0.9\linewidth]{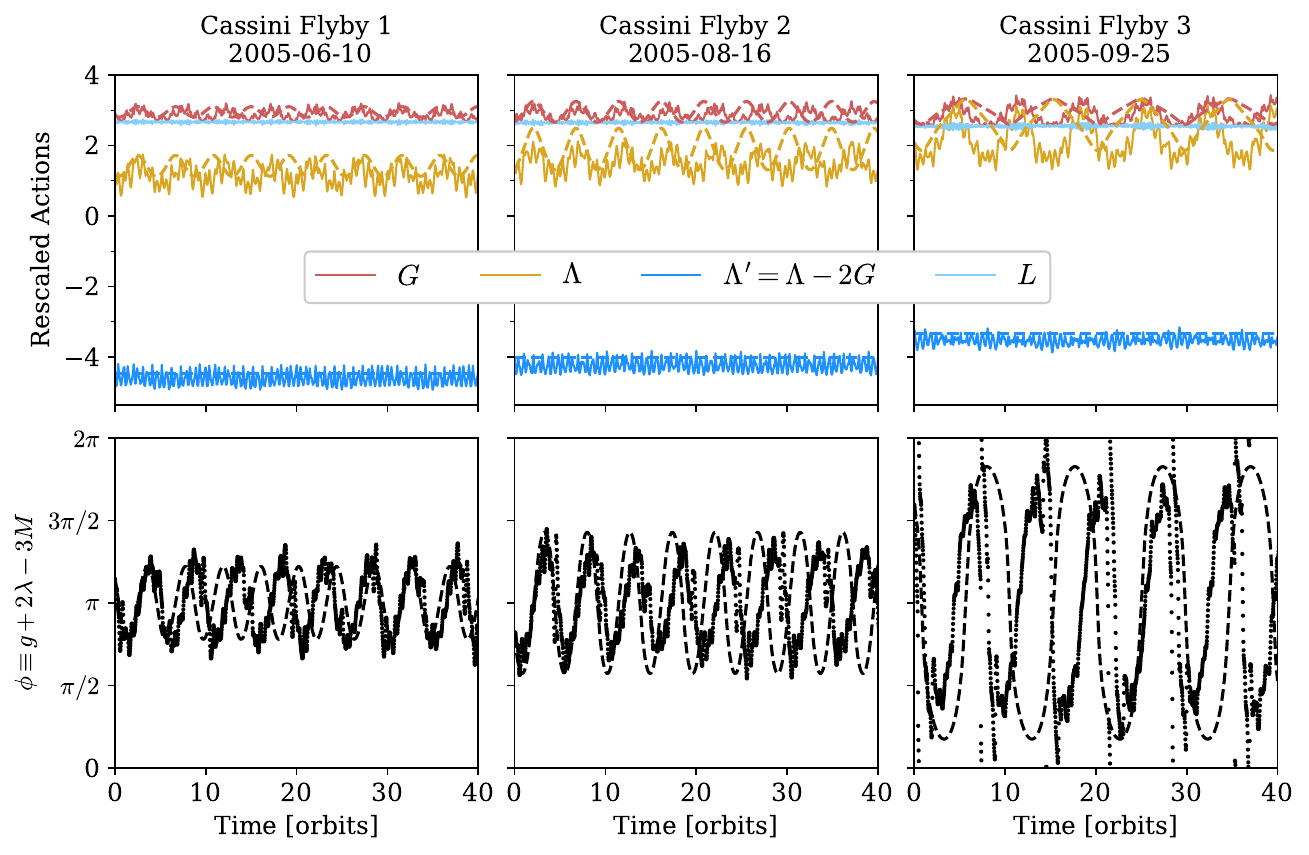}
    \caption{Same Cassini flybys as Fig.~\ref{fig:actions}, except that the initial conditions have been modified to incorporate an estimate of the misalignment of Hyperion's principal rotation axes relative to the axes of the best-fitting ellipsoid shape model \citep{Harbison2011}.}
    \label{fig:actions_corrected}
\end{figure}

\cite{Harbison2011} attempted to address this problem by allowing the true principal axes to be misaligned relative to the body axes. They found that a significantly better fit to the three Cassini flybys is obtained if the principal axes are rotated from the body axes by the Euler angles $(\theta',\phi',\psi')=(40^\circ,20^\circ,10^\circ)$. The recomputed rotation states with this additional transformation are given in Tables \ref{tab:cassini} and \ref{tab:andoyer}. We again integrated the full Hamiltonian $\mathcal{H}$ and the integrable resonant Hamiltonian $\mathcal{H}_3^1$ starting from these initial conditions (Fig.~\ref{fig:actions_corrected}). 

With these corrections, the first two Cassini flybys show clear librating behavior of $\phi$, while the final one indicates circulation very close to the separatrix. Although speculative, this evidence suggests that Hyperion may in fact be inside the 3:1 nutation-orbit resonance! If true, it most likely evolved into this state because of the tidal stresses of NPA rotation, which is quickly damped \citep{Burns1973}.

\subsection{Chaos from resonance overlap}
We have shown that an integrable single resonance model of Hyperion's rotational dynamics reproduces much of its short term behavior. However, in reality even $\hat{\mathcal{H}}^1$ contains an infinite number of resonances and is not integrable. Chaos is expected to arise when the unperturbed resonances would overlap in phase space \citep{Chirikov1979}. The width of each resonance grows with $\rho$ and, except in the case of the 2:1 resonance, with eccentricity. Therefore, at sufficient eccentricity and asphericity of Hyperion, we expect that nutation-orbit coupling will be chaotic, in close analogy to the chaotic spin-orbit evolution originally discovered by \cite{Wisdom1984}.

\begin{figure}
    \centering
    \includegraphics[width=0.5\linewidth]{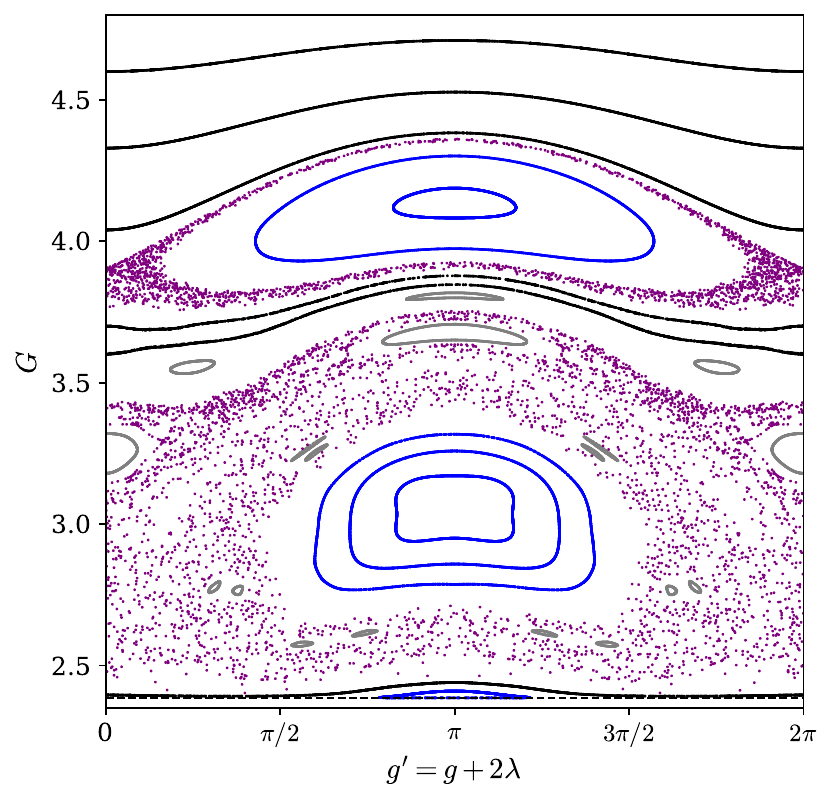}
    \caption{Surface of section of Hamiltonian $\hat{\mathcal{H}}^1$ using the $L$ and $\Lambda'$ values of Cassini Flyby 2. Librational solutions around the main nutation-orbit resonances are plotted in blue, while secondary resonances are shown in gray. A large chaotic sea (purple) surrounds the 3:1 resonance; a smaller chaotic separatrix surrounds the 4:1. At rapid spin rates, many invariant tori are preserved (black).}
    \label{fig:chaossection}
\end{figure}

To study this chaotic behavior, we construct a Poincaré surface of section of $\hat{\mathcal{H}}^1$ by sampling $g'(t), G(t)$ at pericenter, that is, when $t$ is a multiple of $2\pi$, for a variety of initial conditions.
Fig.~\ref{fig:chaossection} shows this surface of section for the values of $e$, $\Lambda'$, and $L$ taken from the 2005-09-25 flyby (neglecting possible misalignment of the principal axes with the body long axes). The 2:1, 3:1, and 4:1 nutation-orbit resonant islands are clearly visible, although the 2:1 resonance is nearly coincident with the boundary of the allowed parameter space at $G=L$. Surrounding the 3:1 resonance is a large chaotic sea, which does not extend fully to the 4:1 resonance. Many secondary resonances, colored in gray, are also visible.

Much of the apparent chaotic behavior in the current rotation of Hyperion can be explained solely by overlap of the $p$:1 nutation orbit resonances that appear in $\hat{\mathcal{H}}^1$. To quantify this chaos, for each of the initial conditions, we estimate the maximum Lyapunov exponent (MLE), $\mathcal{L}$, defined by
\begin{equation}
    \mathcal{L} = \lim_{t\to\infty} \frac{1}{t}\ln \frac{\sqrt{\delta G(t)^2 + \delta g'(t)^2}}{\sqrt{\delta G(0)^2 + \delta g'(0)^2}}.
    \label{eq:lyap}
\end{equation}
Here, $\delta G(t)$ and $\delta g'(t)$ are the solutions to the linearized equations of motion,
\begin{align}
    \dot{\delta G} &= -\frac{\partial^2 \hat{\mathcal{H}}^1}{\partial g' \partial G}\delta G -\frac{\partial^2 \hat{\mathcal{H}}^1}{\partial g' \partial g'}\delta g' \\
    \dot{\delta g'} &= \frac{\partial^2 \hat{\mathcal{H}}^1}{\partial G \partial G}\delta G +\frac{\partial^2 \hat{\mathcal{H}}^1}{\partial G \partial g'}\delta g'
    \label{eq:vareq}
\end{align}
evaluated on the exact trajectory $G(t),g'(t)$ that is the solution to the usual equations of motion \citep{Morbidelli2002}. Of course, the limit in Eq.~\ref{eq:lyap} cannot be exactly computed because the equations must be integrated numerically. Instead, by truncating the limit at a finite $t$, we are actually calculating the finite time MLE (FT-MLE) \citep[e.g.,][]{Benettin1980}. 

\begin{figure}
    \centering
    \includegraphics[width=0.8\linewidth]{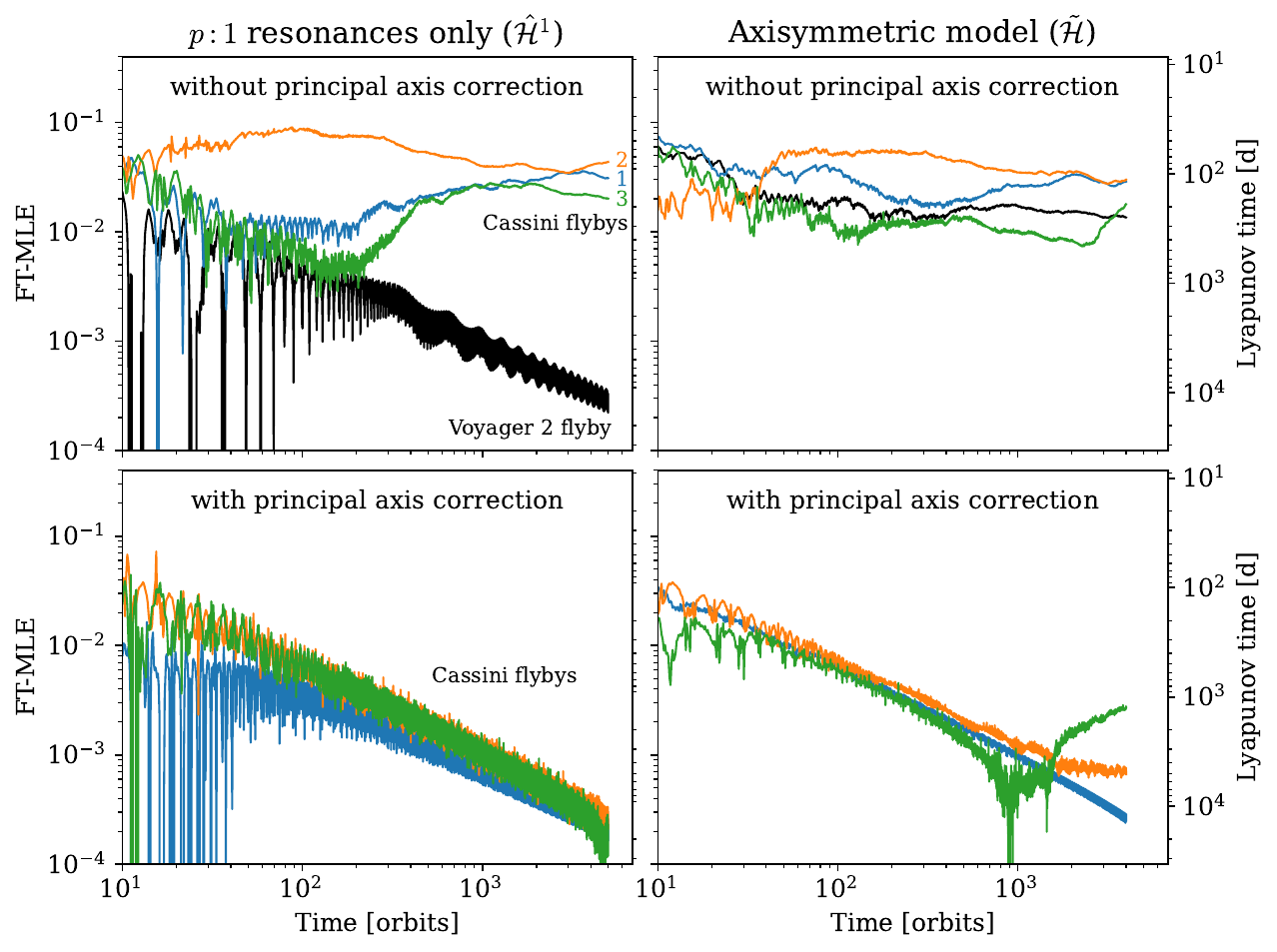}
    \caption{Finite time Maximum Lyapunov exponents (FT-MLEs) for the initial conditions given in Table \ref{tab:andoyer}, integrated under the flow of $\hat{\mathcal{H}}^1$ (left column) and $\tilde{\mathcal{H}}$ (right column). The top row uses the initial conditions reported by \cite{Black1995} and \cite{Harbison2011} without adjusting for misalignment between the apparent body axes and the principal axes; the bottom row incorporates the estimated misalignment from \cite{Harbison2011} for the Cassini flybys. The three Cassini flybys without the misalignment correction clearly have positive FT-MLE in both Hamiltonians, whereas the corrected Cassini flybys are consistent with small or zero FT-MLE. The Voyager 2 flyby is chaotic only under $\tilde{\mathcal{H}}$.}
    \label{fig:lyap}
\end{figure}

Figure \ref{fig:lyap} shows these FT-MLEs for the seven initial conditions from the Voyager 2 and Cassini flybys. We also computed the FT-MLEs for the same seven initial conditions but under the flow of the full axisymmetric Hamiltonian $\tilde{\mathcal{H}}$ and its variational equations, and then modifying Eqs. \ref{eq:lyap} and \ref{eq:vareq} to add the additional coordinates. Positive FT-MLEs are seen for each the Cassini flybys without the principal axis correction, corresponding to a Lyapunov time of 80--150 d. This is consistent with the FT-MLEs found by previous work. \cite{Harbison2011} found Lyapunov times $61.4 \pm 3.6$ d for several initial conditions slightly offset from the observed state, whereas \cite{Black1995} found 66--145 d for a wider range of initial conditions. The Lyapunov times found by \cite{Wisdom1984} are shorter, at less than 40 d. However, their integrations began at much slower rotation rates in order to trigger true chaotic tumbling. Evidently, three degree-of-freedom tumbling is somewhat more chaotic than the overlap of one degree-of-freedom nutation-orbit resonances.

The FT-MLEs under the flow of $\tilde{\mathcal{H}}$ are similar to those of $\hat{\mathcal{H}}^1$ for the Cassini flybys without correcting for principal axes alignment, indicating that most of the chaotic evolution arises from overlap of $p$:1 resonances rather than other effects.The most significant difference appears for Voyager 2 flyby observations, for which the FT-MLE is consistent with zero for $\hat{\mathcal{H}}^1$ but is finite and $\sim 200$ d for $\tilde{\mathcal{H}}$.

Interestingly, Fig.~\ref{fig:lyap} shows that in the initial conditions reported by \cite{Black1995} as well as the principal axis-corrected conditions of \cite{Harbison2011} are not chaotic under $\hat{\mathcal{H}}^1$; that is, their evolution is consistent with zero Lyapunov exponent. These solutions lie within the resonant islands, which remain even after some overlap of neighboring resonances. The corrected Cassini flybys remain consistent with zero or very small Lyapunov exponent under $\tilde{\mathcal{H}}$. It is conceivable that chaos could reappear with the inclusion of non-axisymmetric terms in the Hamiltonian, but the quasi-conservation of $L$ and $\Lambda'$ under $\mathcal{H}$ seen in Figs.~\ref{fig:actions} and \ref{fig:actions_corrected} suggests that this chaos must be relatively weak. Although a precise conclusion is hampered by the lack of a clearly reliable rotational solution, there is a clear hint that the Lyapunov timescale of Hyperion's rotation is considerably longer than originally predicted, and may even approach infinity. We discuss the possible origins of this configuration in Sect.~\ref{sec:discussion}.

\section{Relationship to the barrel instability}
It has been previously noted that there exist spin states which cannot be neatly categorized into planar spin-orbit resonance or chaotic tumbling \citep{Quillen2022}. Relatedly, \cite{Cuk2021} identified the so-called ``barrel instability'' in binary asteroids, in which a prolate satellite rolls slowly along its longest axis while remaining generally aligned with the primary. This state arises in a narrow range of parameter space between synchronous rotation and chaotic tumbling.
In work motivated by small planetary satellites, \cite{Melnikov2020} also found that, for certain parameters, the figure of the satellite was preferentially aligned in the direction of the planet during episodes of quasi-regular rotation even while rotation appeared to be chaotic. 

These phenomena arise naturally from nutation-orbit resonances. 
During chaotic evolution, the system will sometimes diffuse into the resonant domain and remain there for an extended duration. 
In such a configuration, the actions will evolve quasi-periodically and the satellite will be preferentially aligned towards the planet.
If dissipative effects are included, as in \cite{Cuk2021}, the capture into resonance can be permanent.

\subsection{Single harmonic}
The effects of preferred orientation originate from the resonances in the $\cos(2g+2\lambda-2f)$ harmonic of $\tilde{\mathcal{H}}$. Repeating the procedure of Sect.~\ref{sec:sing} with only this harmonic, we have
\begin{align}
    \hat{\mathcal{H}}^2 = \frac{G^2}{2} + \frac{\rho}{16r^3}\left( - 2\frac{L^2}{G^2} - 2\left(1 + \frac{\Lambda'}{G}\right)^2 + 6\frac{L^2}{G^2} \left(1 + \frac{\Lambda'}{G}\right)^2\right. \nonumber \\
    \left. + \left(1 - \frac{L^2}{G^2}\right)\left(2+\frac{\Lambda'}{G}\right)^2\cos(2g' - 2f) \right)
    \label{eq:H11}
\end{align}
where in this case the canonical transformation is $g'=g+\lambda$, $\Lambda'=\Lambda-G$, $G'=G$ and $\lambda'=\lambda$, and $\Lambda'$ is conserved.
Then, expanding in powers of $e$, and again averaging over terms that depend only on $M$, we obtain
\begin{align}
    \hat{\mathcal{H}}^2 \approx \frac{G^2}{2} + \frac{\rho}{16}\left(\left(1 + \frac{3e^2}{2}\right)\left(- 2\frac{L^2}{G^2} - 2\left(1 + \frac{\Lambda'}{G}\right)^2 + 6\frac{L^2}{G^2} \left(1 + \frac{\Lambda'}{G}\right)^2\right)\right. \nonumber\\
    + \left(1 - \frac{L^2}{G^2}\right)\left(2+\frac{\Lambda'}{G}\right)^2\left(\left(1-\frac{5e^2}{2}\right)\cos(2g' - 2M) \right. \nonumber\\ 
    + \frac{7e}{2}\cos(2g'-3M)  \nonumber\\
    -\frac{e}{2}\cos(2g'-M) \nonumber\\
    + \left.\left.\frac{17e^2}{2}\cos(2g'-4M) \right)\right) + O(e^3).
    \label{eq:H1d}
\end{align}
Clearly, resonances occur when $2\dot{g}' = 2\dot{g} + 2\dot{\lambda} \simeq p$ where $p$ is a nonzero integer, and again the leading term of the coefficient of each harmonic scales as $e^{|p-2|}$. We refer to these as $p$:2 nutation-orbit resonances, to distinguish them from the resonances studied in Sect.~\ref{sec:sing}. While these resonances also involve commensurability between the nutation and orbit frequencies, the quantitative dependence on the actions, which sets the resonance widths and libration frequencies, differs significantly from the first type. In particular, these resonances do not disappear as $L\rightarrow 0$, $\Lambda\rightarrow G$, but instead approach the standard planar spin-orbit resonances.

\cite{Cuk2021} and \cite{Quillen2022} described the barrel instability as related to, or ``embedded,'' in the synchronous resonance. The appropriate strategy is thus to select only the $2g'-2M$ term to obtain an integrable Hamiltonian, in effect averaging over the other harmonics. Extending the phase space as before and then performing the canonical transformation $\phi=g'-M$ and $\Theta=G+T$, we obtain
\begin{align}
    \mathcal{H}_2^2 = \frac{G^2}{2} - G + \frac{\rho}{16}\left(\left(1 + \frac{3e^2}{2}\right)\left(2\left(1 - 3\frac{L^2}{G^2}\right)\left(\frac{1}{3} - \left(1 + \frac{\Lambda'}{G}\right)^2\right)\right)\right. \nonumber \\
    + \left.\left(1 - \frac{L^2}{G^2}\right)\left(2+\frac{\Lambda'}{G}\right)^2\left(1-\frac{5e^2}{2}\right)\cos(2\phi)\right) + \Theta
\end{align}
which has only a single degree-of-freedom. This form of the expression reveals why a nearly synchronous state with long axis roll is possible: $L/G$ only appears as a second-order correction, and thus $L$ can vary moderately while only marginally changing the position and width of the synchronous resonance.

\begin{figure}
    \centering
    \includegraphics[width=\linewidth]{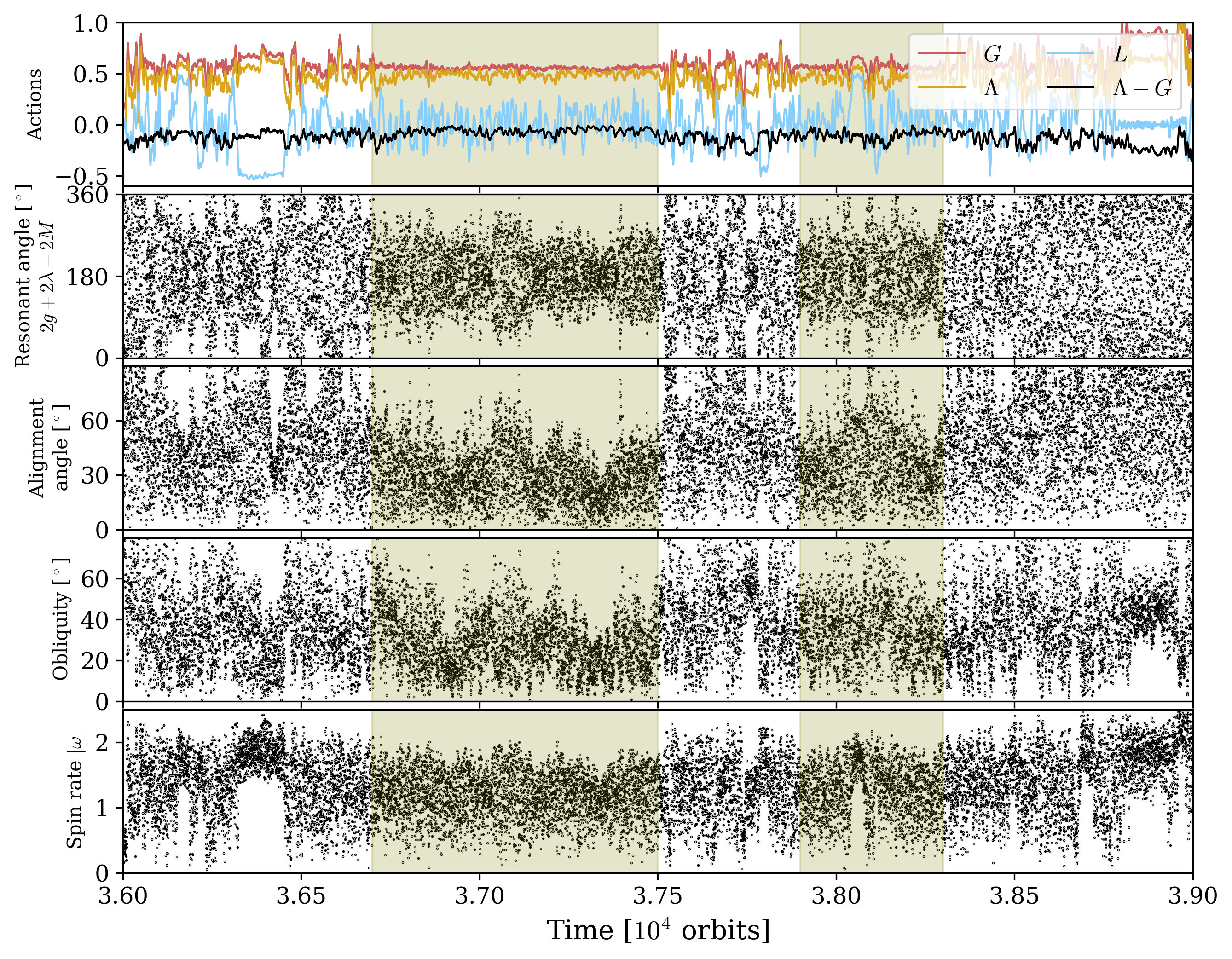}
    \caption{Excerpt from a integration of the full spin Hamiltonian $\mathcal{H}$ for $10^5$ orbits, using the (triaxial) moments of inertia of Hyperion but orbital eccentricity $e=0.01$. The region highlighted in green shows the barrel instability \citep{Cuk2021}, characterized by irregular spin rate (bottom row), obliquity (fourth row), and roll on the long axis (top row, light blue) but a clear preferred orientation of the long axis towards the primary (third row). Analysis in Andoyer coordinates and in the framework of nutation-orbit resonances shows that this behavior is associated with libration of the $2g+2\lambda -2M$ resonant angle (second row).}
    \label{fig:barrel_time}
\end{figure}

As an example, we integrated the full, triaxial, Hamiltonian $\mathcal{H}$ for $10^5$ orbits using Hyperion's moments of inertia and an orbital eccentricity of $e=0.01$, starting from near-synchronous initial conditions. The system quickly begins chaotic tumbling, but also experiences prolonged periods of libration of the $2g+2\lambda-2M$ resonant angle, shown in Fig.~\ref{fig:barrel_time}. At those times, the long axis of the satellite is preferentially oriented towards the primary, while the long axis rolls chaotically, as evidenced by rapid variation in $L$. Considering only the overall spin rate $|\omega|$ and obliquity $\cos^{-1}\Lambda/G$ would suggest that the satellite is chaotically tumbling, but the rotation is in fact quite regular. 

\subsection{Orientation of satellites}
\label{sec:orient}
To quantitatively study the orientation of satellites inside these resonances, suppose that the system lies at the equilibrium point of the $2g+2\lambda-2M$ resonance of $\mathcal{H}_2^2$, so that the actions $G$, $L$, and $\Lambda$ are constant in time and the resonant angle $2\phi \equiv 2g+2\lambda-2M$ remains at its equilibrium point $2\phi_0$. 
Then, the direction cosine describing the orientation of the body $c$ axis towards the planet is (Eq.~\ref{eq:gamma})
\begin{equation}
    \gamma = \sin J \cos(g-\phi_0) \sin g - (\cos J \sin I + \sin J \cos I \cos g)\sin(g-\phi_0) + O(e).
\end{equation}
The time average of $\gamma$ over one period of $g$ is
\begin{equation}
    \langle \gamma \rangle = \frac{1}{2\pi}\int_0^{2\pi} \gamma dg = \sin J \frac{1+\cos I}{2}\sin\phi_0+ O(e).
\end{equation}
For prolate bodies (i.e., when $\rho>0$), $\phi_0=\pi/2$, so $\langle \gamma \rangle$ is nonzero as long as $L \neq G$ and $\Lambda \neq -G$. On the other hand, for oblate bodies such as rapidly spinning bodies in hydrostatic equilibrium, $\rho < 0$ and the stable and unstable equilibria exchange places. In this case, $\phi_0=0$, so that $\langle \gamma \rangle = O(e)$ and there is no strong effect of preferred orientation of the $c$ axis (which in this case is the short axis) to the primary.

Interestingly, this argument does not hold for the 2:1 resonance studied in Sect.~\ref{sec:sing} with resonant angle $\phi \equiv g + 2\lambda - 2M$. In that case, we have
\begin{equation}
    \gamma = \sin J \cos(g/2-\phi_0/2) \sin g - (\cos J \sin I + \sin J \cos I \cos g)\sin(g/2-\phi_0/2) + O(e).
\end{equation}
For the time average, $g$ completes two cycles in one orbit, so we must change the limits of integration to 0 to $4\pi$ and the time average is
\begin{equation}
    \langle \gamma \rangle = \frac{1}{4\pi}\int_0^{4\pi} \gamma dg = O(e).
\end{equation}
In fact, a detailed calculation to second order in eccentricity for arbitrary $p$ (Appendix \ref{sec:A2}) shows that $\langle \gamma \rangle = O(e^2)$ for all $p$:1 resonances. Indeed, Hyperion's long axis is not preferentially oriented towards Saturn in numerical integrations because its dynamics is dominated by the $g + 2\lambda - 3M$, or 3:1, resonance.

\begin{figure}[h]
    \centering
    \includegraphics[width=\linewidth]{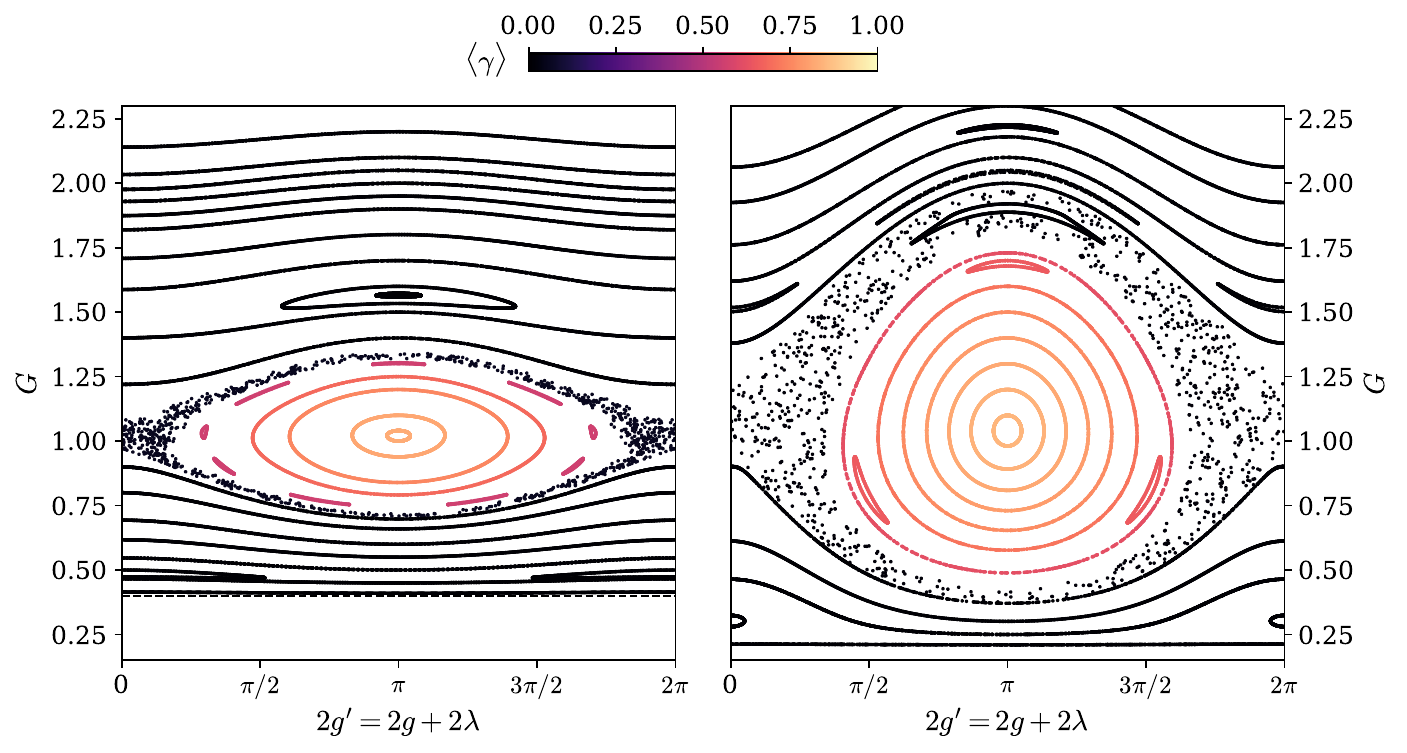}
    \caption{Two Poincaré surfaces of section of $\hat{\mathcal{H}}^2$. In the left panel, we have set $\rho=0.15$, $e=0.01$, $L=0.4$, $\Lambda'=-0.2$. In the right panel, we have set $\rho = 0.9$, $e=0.01$, $L=0.2$, $\Lambda'=-0.3$. Orbits are colored by the time average of $\gamma$, the direction cosine between the long axis and the primary.}
    \label{fig:barrel_section}
\end{figure}

We verified these analytical estimates with numerical integrations of the Hamiltonian. Fig.~\ref{fig:barrel_section} shows two Poincaré surfaces of section for $\hat{\mathcal{H}}^2$, sectioned again at pericenter. In both cases we used $e=0.01$. In the left panel, the satellite is only slightly aspherical, and has a significant long axis spin and out-of-plane rotation. The 1:2, 2:2, and 3:2 resonances are clearly visible and separated. Only the 2:2 resonance shows preferred orientation of the long axis towards the primary. The strength of the preferred orientation effect decreases with increasing libration amplitude. In the right panel, the asphericity is much larger, comparable to that of Hyperion, causing the resonances to be much larger, and the out-of-plane angular momentum is larger than the long axis angular momentum. The 2:2 and 3:2 resonances overlap to create a large chaotic sea, but the 2:2 resonant island remains. Only trajectories in the 2:2 resonant island are preferentially oriented towards the primary.

\section{Diffusion from multiple harmonics}
\label{sec:diffusion}

We now turn to the difficult task of studying $\tilde{\mathcal{H}}$ without neglecting any harmonics. We do not aim to capture the full range of its behavior, but we focus on the parameters of Hyperion in an attempt to understand some of the qualitative features. We integrated $\tilde{\mathcal{H}}$ for $10^5$ orbits using the moments of inertia from \cite{Harbison2011} and $e=0.1$. The only conserved action under $\tilde{\mathcal{H}}$, $L$, is taken from the 2005-06-10 flyby, while the other actions are initialized so that the satellite begins in purely long-axis rotation with $1^\circ$ obliquity.

\begin{figure}
    \centering
    \includegraphics[width=0.95\linewidth]{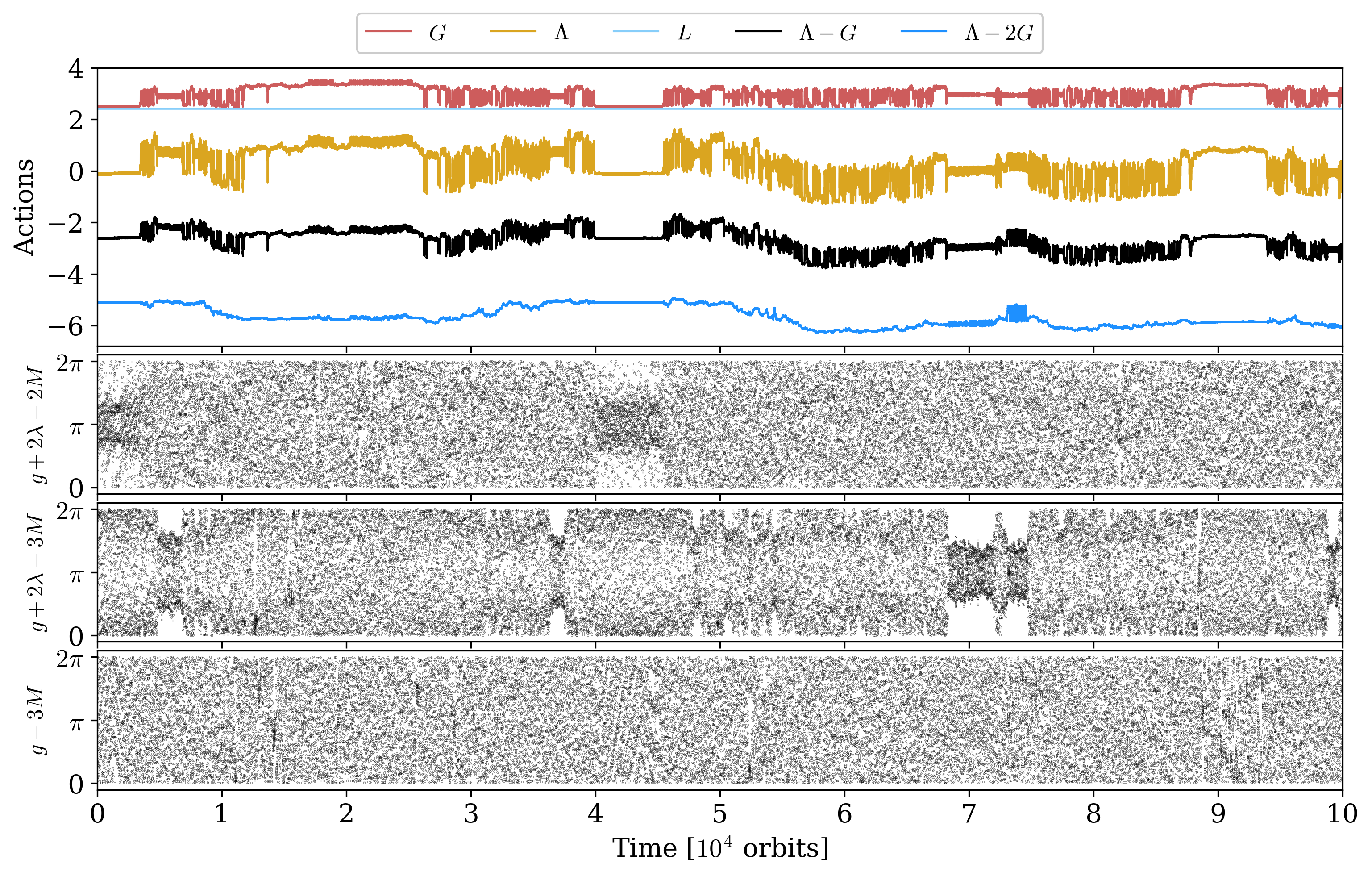}
    \caption{Evolution of the actions and three resonant angles under the flow of the axisymmetric but otherwise complete Hamiltonian $\tilde{\mathcal{H}}$. The action $L$ is perfectly conserved, while the quasi-integral $\Lambda-2G$ varies slowly over thousands of orbits. Temporary capture into resonance occurs frequently, but outside of resonance $G$ and $\Lambda$ evolve rapidly and chaotically.}
    \label{fig:long}
\end{figure}

Figure \ref{fig:long} shows the actions over the integration. We have also plotted the three resonant angles that show clear episodes of libration, $g+2\lambda-2M$, $g+2\lambda -3M$, and $g-3M$. The most notable feature is that the actions $G$ and $\Lambda$ experience rapid and chaotic oscillations on the timescale of tens of orbits, the same order of magnitude of the libration frequency of nutation-orbit resonances. However, the quantity $\Lambda-2G$, which is conserved in the integrable model of Hyperion's rotation $\mathcal{H}_3^1$, diffuses slowly through the integration and does not undergo rapid oscillations except for a short duration around $t=7\times 10^4$. Large changes in the obliquity and total angular momentum of the satellite can occur with minimal change in $\Lambda-2G$, such as near $t=0.5\times 10^4$ and between $t=8\times 10^4$ and $t=10^5$. The $g+2\lambda-2M$ and $g+2\lambda -3M$ resonant angles also move in and out of libration. Episodes of libration typically correspond to relatively small oscillations in the actions, similar to the current observed state of Hyperion. At least one stretch of rapid evolution of $\Lambda-2G$ around $t=5.3\times 10^4$ appears to be related to the near-libration of the $g-3M$ resonant angle.

The origin of this behavior is likely to be a hierarchy in the strength of resonant terms in $\tilde{\mathcal{H}}$ that do and do not conserve $\Lambda-2G$. The most important terms depend on $g$ and $\lambda$ only through the combination $g+2\lambda$, creating a symmetry that conserves $\Lambda-2G$ over shorter timescales. Slow harmonics that break this symmetry -- because they contain other linear combinations of $g$ and $\lambda$ -- appear at higher orders in $\rho$ and/or $e$ and thus their widths are smaller. These harmonics can change $\Lambda - 2G$, but only on much longer timescales as a consequence of their small coefficients. Therefore, for this set of parameters, we can refer to $\Lambda-2G$ as a quasi-integral of motion \citep{Mogavero2023}. 

Furthermore, \cite{Mogavero2023} demonstrated that in the secular Solar System problem, the presence both of actions displaying rapid chaos as well as slowly evolving quasi-integrals is reflected in a large timescale separation in the spectrum of all Lyapunov exponents. Although there is not an exact one-to-one correspondence, in some cases specific Lyapunov exponents can be related to the widths of the strongest resonances able to break a particular symmetry (i.e., modify a certain action).

This can be understood more quantitatively using the perturbative Lie series expansion of the Hamiltonian to remove fast harmonics. The ``averaging'' we performed in Sect.~\ref{sec:sing} to neglect all but the sole slow $g+2\lambda - 3M$ harmonic, which appeared at order $e^1 \rho^1$, actually introduces new harmonics at order $e^1 \rho^2$, some of which may also be slow. Specifically, we decompose $\tilde{\mathcal{H}} = \mathcal{H}_0 + \rho \mathcal{H}_1$ where $\mathcal{H}_1$ contains the terms over which we wish to average. Then, the first step of perturbative expansion is \citep[e.g., ][]{Morbidelli2002}
\begin{equation}
    \mathbf{H} = \mathcal{H}_0 + \rho \mathcal{H}_1 + \rho \{\mathcal{H}_0, \chi\} + \rho^2 \{\mathcal{H}_1,\chi\} + \frac{\rho^2}{2}\{\{\mathcal{H}_0,\chi\},\chi\} + O(\rho^3)
\end{equation}
where $\chi$ is a generating Hamiltonian chosen such that the homologic equation
\begin{equation}
    \mathcal{H}_1 + \{\mathcal{H}_0,\chi\} = \overline{\mathcal{H}}_1
\end{equation}
is satisfied by some $\overline{\mathcal{H}}_1$ that depends only on the actions (i.e., averaged).\footnote{A careful reader might be concerned that this series is not obviously convergent because $\rho>1$ for Hyperion. However, the proper calculation involves a rescaling of terms so that $\mathcal{H}_0$ is of order unity, and in doing so $\rho$ is reduced to $0.1-0.2$.} Here, $\{f,g\}$ is the Poisson bracket. 

In practice, for Hamiltonians such as $\tilde{\mathcal{H}}$ that can be expanded as sums of cosines, the new terms of order $\rho^2$ in $\mathbf{H}$ contain all of the sums and differences of the angles in $\mathcal{H}_1$. The slow terms that arise in this process all have arguments of the form $g+2\lambda-3M$ or $g+4\lambda-3M$. Terms with argument $g+2\lambda-3M$ still conserve $\Lambda-2G$, while terms with $g+4\lambda-3M$ can modify it. However, we performed an explicit calculation of these terms using the current parameters of Hyperion's rotation state and found that the $g+2\lambda-3M$ terms at order $e^1 \rho^2$ have coefficients approximately one order of magnitude larger than the coefficients of the $g+4\lambda-3M$ terms. Therefore, diffusion of $\Lambda-2G$ is driven by relatively weak terms, while the bulk of the dynamics conserves it. 

\begin{figure}
    \centering
    \includegraphics[width=0.95\linewidth]{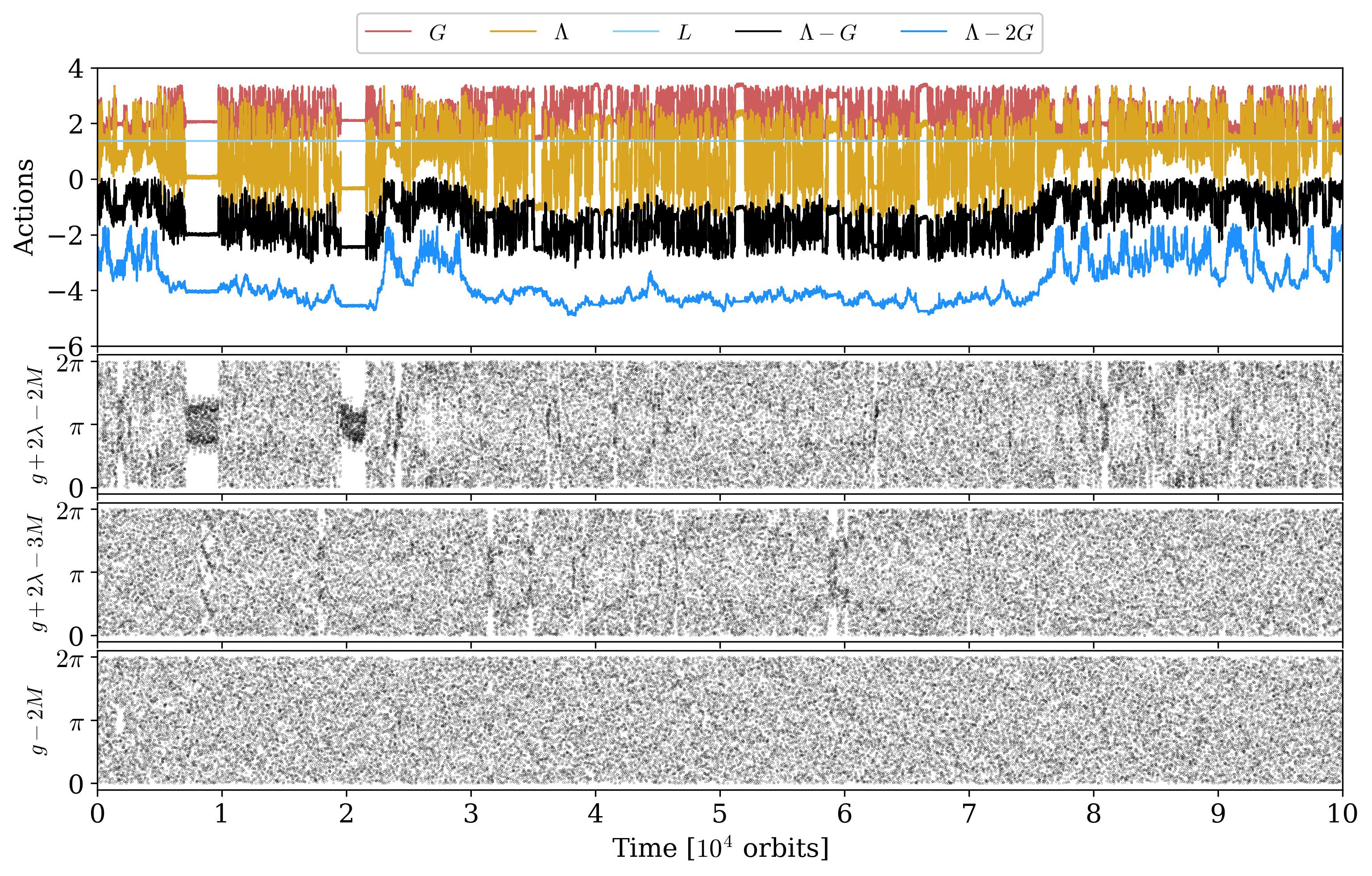}
    \caption{Same as Fig.~\ref{fig:long} but with a lower $L$, which weakens the symmetries that quasi-conserve $\Lambda-2G$.}
    \label{fig:long2}
\end{figure}

Quasi-conservation is not a general property of $\tilde{\mathcal{H}}$ but specific to this choice of parameters. We performed a similar integration as in Fig.~\ref{fig:long} with the same set of parameters except we now set $L=1.4$ (Fig.~\ref{fig:long2}). The lower value of $L$ allows $G$ to reach 2 (recall that $|L|\leq G$), and the system can access a large number of slow terms with arguments $g-2M$, $g+2\lambda -2M$, and $2g+2\lambda-4M$. The coefficients in front of the symmetry-breaking terms are no longer much smaller than those in front of the $g+2\lambda$ terms and $\Lambda-2G$ evolves much faster. In reality, the triaxiality of Hyperion allows $L$ to grow and shrink. Quasi-conservation of $\Lambda-2G$ is a phenomenon that only appears during periods of high $L$. 

\section{Discussion and conclusion}
\label{sec:discussion}
We have investigated the spin-orbit coupling for an axisymmetric satellite in a fixed Keplerian orbit without imposing restrictions on the orientation of its spin axis with an eye towards resolving the discrepancy between observations and simulations of Hyperion's spin dynamics. We find that there is a rich set of nutation-orbit resonances, some of which are clear generalizations of the classic spin-orbit resonances, while others appear only in the presence of non-principal axis (NPA) rotation. Our analysis indicates that Hyperion is not chaotically tumbling, but rather spins rapidly in or near the 3:1 nutation-orbit resonances, whose strength is set by a combination of the asphericity of Hyperion, its orbital eccentricity, and rapid long axis rotation. 

We also show that the barrel instability, a near-synchronous spin state with long axis roll seen in numerical simulations of binary asteroids \citep{Cuk2021}, is actually a capture into the 2:2 nutation-orbit resonance. Under realistic assumptions, out-of-plane rotation and long axis roll can persist and they do not destroy the synchronous resonance.

Furthermore, these resonances can be quite wide for moderate eccentricity and the asphericity typical of asteroids and small planetary satellites. Large chaotic seas, created by resonance overlap, are frequently present. The chaos observed in numerical simulations of Hyperion's rotation can be explained by overlap of these one degree-of-freedom resonances, and two independent quasi-integrals of motion are nearly conserved on timescales of hundreds of orbits. This mechanism is fundamentally different than the overlap of planar spin-orbit resonances proposed by \cite{Wisdom1984}, which only occurs at spin rates less than about twice the orbital frequency. Indeed, at its known spin rate of 4--5 times its orbital frequency, Hyperion cannot be chaotic in the framework of planar spin-orbit resonances because it lies in the regime of preserved KAM tori (and above the top of Fig.~2 of \cite{Wisdom1984}). More detailed observations are needed to precisely constrain the rotation state of Hyperion and determine the degree of chaos present, if any.

We also studied the long term evolution of the unaveraged axisymmetric Hamiltonian. Over hundreds to tens of thousands of orbits, the system alternates between regimes of rapid chaos and highly regular behavior, the latter of which is often accompanied by temporary capture into resonance. Furthermore, for the specific parameters of Hyperion, the strongest available resonances conserve $\Lambda-2G$, resulting in a strong separation of timescales: rapid chaos from overlap of nearby nutation-orbit occurs on top of slow diffusion of the quasi-integral $\Lambda-2G$.

We wish to emphasize that to an extent, the conflation of quasi-regular and fully chaotic states in the literature arises from examination of only the angular velocities and orientation angles of the spin. Instead, in many cases, the Andoyer actions, which have units of angular momentum, have much more regular behavior. For example, in $\hat{\mathcal{H}}^2$, which is the direct generalization of planar spin-orbit resonances, the obliquity $I$ is not conserved. However, the related quantity $\Lambda'=G(\cos I - 1)$, which also expresses the magnitude of rotation away from the orbital plane, is conserved. In ``secular'' regimes where the fast rotation angles can be averaged out, $G$ itself is conserved and the distinction disappears. 

Our work can be extended in several ways. The assumption of axisymmetry can be relaxed to properly model, for example, chaotic tumbling and the long axis spin up of Hyperion seen originally in the simulations of \cite{Black1995}. In contrast to the seven harmonics of $\tilde{\mathcal{H}}$ (each of which is in fact an infinite series in $e$), the triaxial Hamiltonian $\mathcal{H}$ has 23 harmonics. It is conceivable that some approximations could be applied to obtain estimates on qualitative features such as the onset of chaotic tumbling, especially in the low eccentricity or nearly-spherical limit. We have also ignored the coupling of the satellite spin onto its orbit, an effect which is especially important for asteroid binaries \citep{Fahnestock2008}. In principle this could be included by adding additional Keplerian terms to the Hamiltonian \cite[e.g.,][]{Boue2009}.

The most important effects that we have neglected are the impacts of non-conservative forces, particularly tidal dissipation and radiative forces. In reality, any body whose spin is not aligned to its axis of largest moment of inertia experiences internal stresses and dissipates energy \citep{Efroimsky2000}. The rate of dissipation depends sensitively on the spin rate and nutation angle $J$, but can be quite rapid \citep{Burns1973,Frouard2018,Quillen2019}. It is not clear whether the simplest prescriptions of wobble damping used in numerical simulations, in which a force always acts to push the body to rotate on its axis of maximum inertia \citep[e.g.,][]{Meyer2023}, are applicable in this case. Viscoelastic simulations of \cite{Quillen2022} -- which appear to show capture into nutation-orbit resonances -- suggest that NPA rotation can be long-lived following capture. Radiative forces are also crucial as they provide another source of angular momentum that can change both the spin state and the orbit of the satellite through the BYORP effect \citep{Cuk2005}. While comprehensive study is necessary to evaluate the role of tidal and radiative forces in both capture and escape from these resonances, the work carried out in this article provides a Hamiltonian framework, within which more sophisticated models can be developed.

Finally, tidal dissipation may be the key to understanding the origin of Hyperion's unexpected rotation state. Long-axis rotation is the maximum energy configuration, assuming fixed angular momentum, and thus dissipation should cause it to rotate on the shortest axis. However, resonant fixed points generically act as attractors in dissipative dynamical systems. The precise interaction between these two competing effects is not clear. Hyperion's case is complicated by the fact that its orbital eccentricity varies on two timescales due to the mean-motion resonance with Titan: a two year oscillation over the resonant libration cycle \citep{Duriez1997}, and a slow growth from Titan's outward migration \citep{Colombo1974}. Conversely, Hyperion's own rapid NPA rotation can act to reduce its eccentricity \citep{Goldberg2024}. These changes in eccentricity manifest as changes to the rotation state: regular regions can merge into the chaotic sea at high eccentricity, and vice versa. One possibility is that tidal dissipation drives Hyperion into a regular or quasi-regular rotation state at which point dissipation becomes small. Then, its eccentricity eventually grows large enough that the resonant island it sits in disappears, and it is forced into chaotic tumbling. The resulting NPA rotation damps the eccentricity and eventually allows capture into a resonant island. Proper investigation of this scenario would require treatment of orbit-to-spin coupling (the subject of this work), spin-to-orbit coupling, and spin-dependent tidal dissipation. 

\begin{acknowledgements}
We are grateful to the referee for improving the manuscript. We thank Yubo Su, Gabriele Pichierri, Alessandro Morbidelli, Antoine Petit, Phil Nicholson, and Federico Mogavero for insightful discussions. This work was funded by the ERC project N. 101019380 ``HolyEarth.'' K.~B. thanks the Packard Foundation, the Caltech Center for Comparative Planetary Evolution (3CPE), and the National Science Foundation (Grant AST 2109276) for their support. 
\end{acknowledgements}

\bibliographystyle{aa}
\bibliography{main}% common bib file

\begin{appendix}

\section{Coordinate transformations}\label{sec:A1}
In this section we give a description of the reference frames and coordinate systems we use and how to transform between them. 

We use two sets of orthonormal basis vectors that define two frames. First, the lab frame is an inertial frame defined by basis vectors $\mathbf{\hat{x}}$ along the satellite's eccentricity vector, that is, the pericenter direction, $\mathbf{\hat{y}}$ parallel to primary's spin, and $\mathbf{\hat{z}}$ chosen to make an orthogonal right-handed coordinate system. We also define the body frame by basis vectors $\mathbf{\hat{a}}$, $\mathbf{\hat{b}}$, $\mathbf{\hat{c}}$ which are the body's $a$-, $b$-, and $c$-axes.

For a given vector $\mathbf{x}$, we will use the notation $\mathbf{x}_\textrm{lab}=(p,q,r)$ to describe the coordinates in the lab frame, that is, $\mathbf{x}=p \hat{\mathbf{x}} + q \hat{\mathbf{y}} + r \hat{\mathbf{z}}$. Similarly, $\mathbf{x}_\textrm{body}=(\rho,\sigma,\tau)$ means $\mathbf{x}=\rho \hat{\mathbf{a}} + \sigma \hat{\mathbf{b}} + \tau \hat{\mathbf{c}}$. 

Applying this to the spin vector of satellite, we define $\pmb{\omega}_\text{body} = (\omega_a, \omega_b, \omega_c)$ and $\pmb{\omega}_\textrm{lab}=(\omega_x,\omega_y,\omega_z)$. Then, 
\begin{equation}
    \pmb{\omega}_\textrm{lab} = R_\textrm{body\textrightarrow lab} \pmb{\omega}_\textrm{body}
\end{equation}
where $R_\textrm{body\textrightarrow lab}$ is the operator rotating the coordinates from the body frame to the lab frame. 
We can express $R_\textrm{body\textrightarrow lab}$ as a standard rotation matrix by Euler angles, $R(\theta,\phi,\psi)$, using the intrinsic rotation ZXZ convention. These Euler angles are used, for example, in Table \ref{tab:cassini}.

Andoyer canonical coordinates require the use of an intermediate frame aligned to the spin angular momentum which we denote ``And.'' Specifically, the $z$-axis of this frame is parallel to the angular momentum vector $\mathbf{G}$, and the $x$-axis lies on the intersection of the plane normal to $\mathbf{G}$ and the lab $xy$-plane. Finally, the $y$-axis is chosen to define a right-handed coordinate system.

The transformation from body coordinates to lab coordinates can then be decomposed into transformations from the body frame to the Andoyer frame, and then from the Andoyer frame to the lab frame,
\begin{equation}
    R_\textrm{body\textrightarrow lab} = R_\textrm{And\textrightarrow lab} R_\textrm{body\textrightarrow And}.
    \label{eq:rot}
\end{equation}
In Euler angles, these rotations are $R_\textrm{And\textrightarrow lab} = R(\lambda, I,0)$ and $R_\textrm{body\textrightarrow And} = R(g, J, l)$. Recall that $\cos I =\Lambda/G$, and $\cos J = L/G$, and the actions are defined by $G=|\mathbf{G}|$, $L=\mathbf{G}\cdot \hat{\mathbf{c}}$, and  $\Lambda=\mathbf{G}\cdot \hat{\mathbf{z}}$. The Andoyer coordinates are $(G, g, \Lambda, \lambda, L, l)$, which are canonical \citep{Deprit1967}.

These rotations also allow us to express the direction cosines $\alpha$, $\beta$, and $\gamma$ in terms of the Andoyer coordinates. If $\hat{\mathbf{r}}$ is the vector pointing towards the satellite from the primary, we have $\hat{\mathbf{r}}_\textrm{lab}=(\cos f, \sin f,0)$ where $f$ is the true anomaly, and $\hat{\mathbf{r}}_\textrm{body} = (\alpha,\beta, \gamma)$. Then,
\begin{equation}
    \begin{pmatrix}
        \alpha\\
        \beta\\
        \gamma
    \end{pmatrix}=
    R^{-1}_\textrm{body\textrightarrow lab} 
    \begin{pmatrix}
    \cos f\\
    \sin f\\
    0
    \end{pmatrix}
\end{equation}
which gives
\begin{equation}
    \gamma = \sin J \cos(f-\lambda)\sin g - (\cos J \sin I + \sin J \cos I \cos g)\sin(f-\lambda)
\end{equation}
and similar expressions for $\alpha$ and $\beta$ \citep{Lara2010}.

Numerical integrations typically use the Euler rigid body equations to produce body frame spins $\omega_a, \omega_b, \omega_c$ and either Euler angles or quaternions to represent the body orientation. We then calculate the Andoyer coordinates in the following way:
\begin{enumerate}
    \item $G=\sqrt{(A\omega_a)^2 + (B\omega_b)^2 + (C\omega_c)^2}$ and $L=C\omega_c$ are computed directly
    \item The Euler angles or quaternions determine $R_\textrm{body\textrightarrow lab}=R(\theta,\phi,\psi)$.
    \item We transform the angular momentum from body coordinates, $\mathbf{G}_\mathrm{body}=(A\omega_a, B\omega_b, C\omega_c)$, into lab coordinates, $\mathbf{G}_\mathrm{lab}= R_\textrm{body\textrightarrow lab} \mathbf{G}_\mathrm{body}=(\sqrt{G^2 - \Lambda^2} \sin \lambda, -\sqrt{G^2 - \Lambda^2} \cos \lambda, \Lambda)$, then compute $\Lambda$ and $\lambda$, and construct the rotation $R_\textrm{And\textrightarrow lab} = R(\lambda, I,0)$
    \item We invert Eq.~\ref{eq:rot} to obtain $R_\textrm{body\textrightarrow And}= R^{-1}_\textrm{And\textrightarrow lab} R_\textrm{body\textrightarrow lab}$, and extract $g$ and $l$ from $R_\textrm{body\textrightarrow And} = R(g, J, l)$
\end{enumerate}
These calculations are implemented using the \texttt{Rotation} class provided by \texttt{scipy.spatial}.

Two additional transformations are required for the analysis of Hyperion. Firstly, previous work used the convention that $A\leq B\leq C$, so that, for Hyperion, $A < B\approx C$, whereas our convention expects that $A\approx B$. Therefore, to obtain the initial Euler angles and angular velocities (Table \ref{tab:cassini}) and Andoyer coordinates (Table \ref{tab:andoyer}), we must first permute the axes, represented by the rotation $R_\textrm{permute}=R(\pi,\pi/2,\pi/2)$. Secondly, in some cases we apply the correction to the principal axes suggested by \cite{Harbison2011}, which is $R_\textrm{princ}=R(40^\circ,20^\circ,10^\circ)$.
With these additions, Eq.~\ref{eq:rot} now reads
\begin{equation}
    R_\textrm{body\textrightarrow lab} = R_\textrm{princ} R_\textrm{And\textrightarrow lab} R_\textrm{body\textrightarrow And} R_\textrm{permute}.
\end{equation}

\section{Orientation of satellites}\label{sec:A2}
Here we give a more detailed calculation of the average orientation of satellites trapped at the equilibrium point of a nutation-orbit resonance outlined in Sect.~\ref{sec:orient}.
We aim to calculate the time average of the direction cosine of the $c$-axis to the primary,
\begin{equation}
    \gamma(g,\lambda,M) = \sin J \cos(f-\lambda)\sin g - (\cos J \sin I + \sin J \cos I \cos g)\sin(f-\lambda)
\end{equation}
for the two cases of $\phi_0=g+2\lambda - pM$ ($p$:1 nutation-orbit resonance) and $2\phi_0=2g+2\lambda - pM$ ($p$:2 nutation-orbit resonance). We will assume that the evolution of $\lambda$ is slow, that is, we take it to be constant over one orbit. In other words, we calculate
\begin{equation}
    \langle \gamma \rangle_1(p) = \frac{1}{4\pi^2} \int_0^{2\pi} \int_0^{2\pi} \gamma(\phi_0 - 2\lambda + pM, \lambda, M) dM d\lambda
\end{equation}
and 
\begin{equation}
    \langle \gamma \rangle_2(p) = \frac{1}{8\pi^2} \int_0^{2\pi} \int_0^{4\pi} \gamma(\phi_0 - \lambda + pM/2, \lambda, M) dM d\lambda
\end{equation}
where the $4\pi$ arises in the second case because of the presence of terms with $pM/2$.
We carry out the calculations to second order in $e$ and use the standard expansions \citep[e.g.,][]{Murray1999}
\begin{align}
    \cos f = \cos M + e(\cos 2M - 1) + \frac{9e^2}{8}(\cos 3M- \cos M) + O(e^3)\nonumber \\
    \sin f = \sin M + e \sin 2M + e^2 \left(\frac{9}{8}\sin 3M - \frac{7}{8}\sin M\right) + O(e^3).
\end{align}

For the first case, we find that $\langle \gamma \rangle_1(p) = O(e^3)$ for $0 < p \leq 4$, and thus alignment does not occur in $p$:1 nutation-orbit resonances. In the second case for $p$:2 resonances, we find that
\begin{align}
    \langle \gamma \rangle_2(1) &= O(e^3)\nonumber\\
    \langle \gamma \rangle_2(2) &= \left(1 - e^2\right) \sin J \frac{1+\cos I}{2} \sin \phi_0 + O(e^3)\nonumber\\
    \langle \gamma \rangle_2(3) &= O(e^3)\nonumber\\
    \langle \gamma \rangle_2(4) &= e \sin J \frac{1+\cos I}{2} \sin \phi_0 + O(e^3)
\end{align}
and thus alignment only occurs for even values of $p$ but with a coefficient of $e^{(p-2)/2}$. 
These represent idealized scenarios where the amplitude of libration around the fixed point is zero. In reality, finite libration amplitude weakens the alignment effect somewhat, and thus apparent alignment is unlikely to be noticed apart from the $p=2$ resonance. 

\end{appendix}

\label{LastPage}
\end{document}